\documentclass[prb,showpacs,twocolumn,floatfix]{revtex4}
\usepackage{graphicx}
\begin{document}
\def\rhov{{\mbox{\boldmath{$\rho$}}}}
\def\tauv{{\mbox{\boldmath{$\tau$}}}}
\def\Lambdav{{\mbox{\boldmath{$\Lambda$}}}}
\def\sigmav{\sigma}
\def\xiv{{\mbox{\boldmath{$\xi$}}}}
\def\chiv{{\mbox{\boldmath{$\chi$}}}}
\def\rhov{{\mbox{\boldmath{$\rho$}}}}
\def\phiv{{\mbox{\boldmath{$\phi$}}}}
\def\piv{{\mbox{\boldmath{$\pi$}}}}
\def\psiv{{\mbox{\boldmath{$\psi$}}}}
\def\oh{{\scriptsize 1 \over \scriptsize 2}}
\def\ot{{\scriptsize 1 \over \scriptsize 3}}
\def\of{{\scriptsize 1 \over \scriptsize 4}}
\def\tf{{\scriptsize 3 \over \scriptsize 4}}
\title{Effect of Inversion Symmetry on Incommensurate Order
in Multiferroic RMn$_2$O$_5$, R=rare earth.}

\author{A. B. Harris[1], M. Kenzelmann[2], Amnon Aharony[3], and
O. Entin-Wohlman[3]}

\affiliation{ [1]Department of Physics and Astronomy,
University of Pennsylvania, Philadelphia, PA 19104}
\affiliation{ [2]Laboratory for Solid State Physics, ETH Zurich, CH-8093
Zurich, Switzerland and Laboratory for Neutron Scattering, ETHZ \& PSI,
CH-5232 Villigen PSI, Switzerland}
\affiliation{ [3]Department of Physics and the Ilse Katz Center for
Meso-and Nano-Scale Science and Technology, Ben Gurion University,
Beer Sheva 84105 ISRAEL}
%%% ----------------------------------------------------------------------
\date{\today}

\begin{abstract}
Starting from the irreducible representations of the group of
the wave vector we construct the spin
wave functions consistent with inversion symmetry, neglected in the
usual representation analysis.  We obtain the relation between the
basis functions of different members of the star of the wave vector.
We introduce order parameters and determine their transformation
properties under the operations of the space group of the paramagnetic
crystal.  The results are applied to construct terms in the
magnetoelectric interaction which are quadratic and quartic in the
magnetic order parameters.  The higher order magnetoelectric interactions
can in principle induce components of the spontaneous polarization
which are not allowed by the lowest order magnetoelectric interaction.
We also obtain the relation between the spin wave functions of the
incommensurate phase and those of the commensurate phase which lead to
analogous relations between the order parameters of these two phases.
\end{abstract}
\maketitle

\section{INTRODUCTION}

The problem of determining the symmetry of incommensurate (IC) magnetic
order from diffraction experiments is an old one and is the subject
of several well-known reviews.\cite{BERTAUT,ROSSAT} 
The reviews are based on the idea that the spin structure that develops at
a continuous transition must transform
like an irreducible representation (irrep) of the group of
operations which leaves the IC wave vector ${\bf q}$
invariant.\cite{LL} However, perhaps surprisingly, these
standard references do not exploit additional restrictions that are
due to inversion symmetry when that operation is not a member
of the group of the wave vector.  Although the group theoretical
formalism for doing this has been described\cite{VILL,RC07} and
these restrictions had previously been used to aid in structure
determinations,\cite{NVO,TMO,PRB,ABH} the effect of inversion symmetry
is often not included in the classification of possible magnetic
structures.  

Here we perform the requisite analysis for the star of wave vectors
of the IC phases\cite{LTIC,SK05,SK04a}
of the ``125" systems, RMn$_2$O$_5$, where R is a rare
earth ion, which may be magnetic or not ({\it e. g.} when R
is yttrium).  The interest in these materials stems from the
fact that they exhibit ferroelectricity\cite{KS95,AI96,IK02,IK03}
whose onset coincides with a magnetic ordering 
transition.\cite{SK04b,HK05,SK04c,DH04,DH05}
We show that when inversion symmetry is
taken into account, there are about half as many degrees of freedom
that describe the basis functions of the irreducible representations compared
to an analysis when inversion symmetry is overlooked.  Even when an
unrestricted fit (not taking account of any symmetry) is
performed,\cite{HK07} it is useful to have the results
of the present paper to see if the hypothesis of a single irrep\cite{FN}
holds.  Thus, it is clear that magnetic structure determination
using an approach that includes inversion symmetry will
lead to an increase in the accuracy of the structure determinations.
Finally, this approach leads naturally to the introduction of
order parameters which have symmetry properties that we explicitly
display and in terms of which a Landau expansion was developed
for a number of systems\cite{NVO,TMO,PRB,ABH} and which has led
to a generic magnetoelectric (ME) phase diagram for the 125's.\cite{HAE}
The purpose of the present paper is to a) analyze the symmetry of the
various IC phases, b) show how the symmetry implies relations
between order parameters of different symmetry magnetic phases,
and c) analyze the symmetry of the ME interactions
which explain the appearance of ferroelectric order at some
of the magnetic phase transitions.

\begin{figure}[ht]
\begin{center}
\includegraphics[width=8.0 cm]{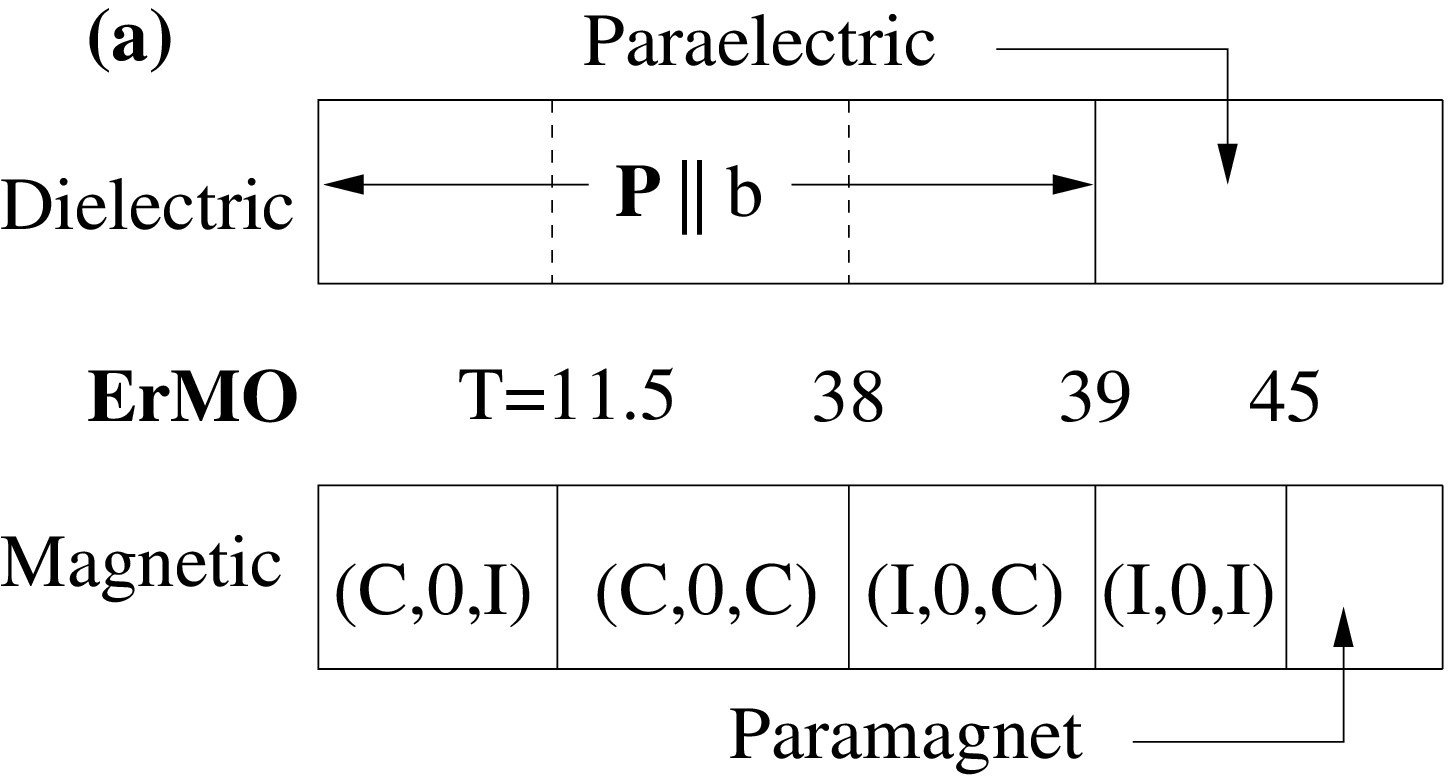}

\vspace{0.2 in}
\includegraphics[width=8.0 cm]{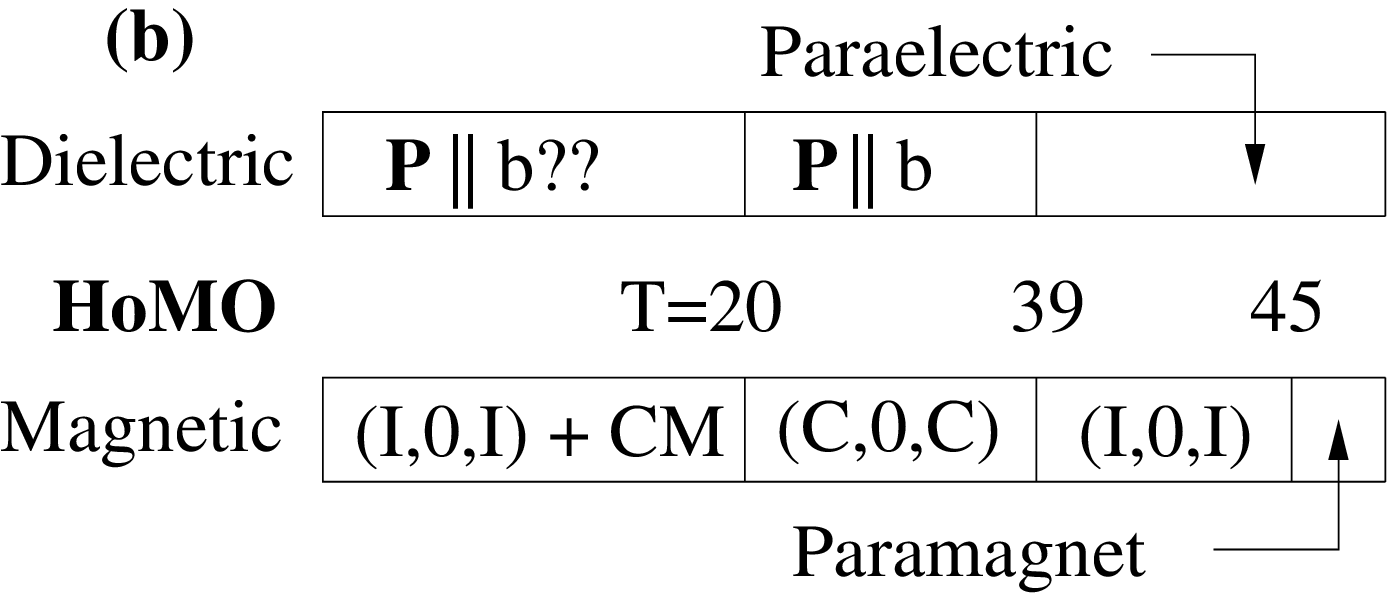}
\caption{\label{ERMO}
(a): The ME phase diagram of 
ErMn$_2$O$_5$.\protect{\cite{SK04b}}
Here $(X,0,Z)$ indicates the nature of the wave vector.
If $X=C$ ($Z=C$), then $q_x=1/2$ $(q_z=1/4)$.  If $X=I$
$(Z=I)$, then $q_x$ ($q_z$) is IC, but close to 1/2 (1/4).
The dashed lines indicate temperatures at which an anomaly in the
$b$-component of the dielectric constant was observed.
$P || b$ indicates that the system has a spontaneous polarization
aligned along ${\bf b}$ (for $T<39$ K).
(b): Same for HoMn$_2$O$_5$.\protect{\cite{HK05}}  For $T>39$ K, $q_z<1/4$
and for $T<20$ K, the $(I,0,I)$ phase has $q_z> 1/4$ and the system is either
paraelectric or weakly ferroelectric.}
\end{center}
\end{figure}

Briefly, this paper is organized as follows.  In Sec. II we list the
results obtained using the canned program MODY for the IC phase
and we show how to modify this to take account of inversion symmetry.
Here order parameters are introduced as the complex amplitudes
of the spin wave functions.
In Sec. III we show how, having obtained the basis functions for one
member of the star of ${\bf q}$, one can determine the basis functions
for all the other wave vectors in the star of ${\bf q}$. Here we also
determine how the order parameters transform under all the operations
of the space group.  Having determined the symmetry properties of
the order parameters we are able, in Sec. IV, to construct the
lowest order (trilinear) ME interaction which explains the orientation
of the observed magnetically induced spontaneous polarization.  Here we
show that higher order and {\it Umklapp} ME interactions can lead to
small contributions to all components of the spontaneous polarization.
In Sec. V we discuss how the basis functions in the IC phase with
$q_x \not= 1/2$ connect to those in the adjacent $q_x=1/2$ phase.
Here we also analyze the symmetry of the special multicritical point for
which $q_x=1/2$.  In Sec. VI we briefly summarize the results of this paper.

\section{CALCULATION}

\subsection{Results without Inversion Symmetry}

The lattice structure of the 125's was determined
by Quezel-Abrunaz {\it et al.}\cite{QUEZ} to be that of the
orthorhombic space group Pbam (\#55 in Ref. \onlinecite{ITC}).
In Table \ref{SPACE} we list the general positions in the
primitive unit cell which define the symmetry operations of
the space group Pbam and in Table II\cite{BUIS1,BUIS2,ALON} we give
the actual positions of the ions for the 125 systems.

\begin{table}
\caption{\label{SPACE} General positions within the unit cell
for space group Pbam expressed as fractions of the orthorhombic
lattice constants.\protect{\cite{ITC}}
This table defines the space group operations on ${\bf r}=(x,y,z)$.
Here $2_\alpha$ is a two-fold rotation (or screw) about the $\alpha$
axis and $m_{\alpha \beta}$ is a mirror (or glide) $\alpha \beta$
plane.}
\vspace{0.2 in}
\begin{tabular} { || c  | c ||}
\hline
$E{\bf r} \equiv (x,y,z)$ & $2_a {\bf r} \equiv (x+1/2,\overline y + 1/2, 
\overline z)$ \\
$2_b{\bf r} \equiv ( \overline x + 1/2, y+ 1/2, \overline z)$ &
$2_c {\bf r} \equiv (\overline x , \overline y , z)$ \\
${\cal I} {\bf r} \equiv (\overline x, \overline y, \overline z)$
& $m_{bc} {\bf r} \equiv  (\overline x + 1/2, y+1/2, z)$ \\
$m_{ac} {\bf r} \equiv (x+ 1/2, \overline y+ 1/2, z)$ &
$m_{ab} {\bf r}  = (x,y, \overline z)$ \\
\hline
\end{tabular}
\end{table}

\begin{table}
\caption{\label{SITES} Position $\tauv_n$ (in units of lattice constants)
of the $n$th magnetic ion in the unit cell. (These values are
for HoMn$_2$O$_5$,\protect{\cite{BUIS1,BUIS2}} but are approximately
the same for the other 125's.\protect{\cite{ALON}}) Sites 1-4 are
for Mn$^{3+}$, 5-8 are for Mn$^{4+}$ and 9-12 are for R$^{3+}$ ions.}
\vspace{0.2 in}
\begin{tabular} {|| c | c ||}
\hline
$\tauv_1 = (0.09,0.85,1/2)$ & $\tauv_2 = (0.59,0.65,1/2)$ \\
$\tauv_3 = (0.41,0.35,1/2)$ & $\tauv_4 = (0.91,0.15,1/2)$ \\
$\tauv_5 = (1/2,0,0.25)$ & $\tauv_6 = (0,1/2,0.25)$ \\ 
$\tauv_7 = (0,1/2,0.75)$ & $\tauv_8 = (1/2,0,0.75)$ \\ 
$\tauv_9 = (0.14,0.17,0)$ & $\tauv_{10} = (0.64,0.33,0)$ \\
$\tauv_{11} = (0.36,0.67,0)$ & $\tauv_{12} = (0.86,0.83,0)$\\ 
\hline
\end{tabular}
\end{table}
 
The magnetic and dielectric phases occuring in the 125's are more
complicated and we give a brief overview of them here.
In Fig \ref{ERMO}a and \ref{ERMO}b  we show the ME phase diagrams of
ErMn$_2$O$_5$ (taken from Ref. \onlinecite{SK04b}) and HoMn$_2$O$_5$
(taken from Ref. \onlinecite{HK05})
which exhibit the simultaneous ferroelectric and
magnetic phase transitions.  When cooled from
the paramagnetic phase, the 125's  develop IC order at about 45 K 
in a paraelectric phase described by the wave vectors whose star 
consists of ${\bf q}_\pm = [(1/2-\delta ,0,\pm (1/4 + \epsilon)]$ and
their negatives, where $\delta$ and $\epsilon$ are of order 0.05
or less\cite{SK04b,SK04a,LC06,HK05,GB05,WR05,RE08,SK05,SK04c}
in reciprocal lattice units (rlu's).
Upon further cooling some 125's, such as ErMn$_2$O$_5$ 
(shown in Fig. 1a),\cite{DH05,SK04b}
YMn$_2$O$_5$\cite{AI96,IK03,SK04a,LC06} and TmMn$_2$O$_5$,\cite{SK05,MU98}
exhibit a ferroelectric $(I,0,C)$ phase in which $\epsilon=0$, before
entering a $(C,0,C)$ phase in which $\delta=\epsilon=0$. Other
125's, such as TbMn$_2$O$_5$,\cite{SK04c,AI96,NH04a}
HoMn$_2$O$_5$ (shown in Fig. 1b),\cite{NH04b,HK05,DH05,DH04} and
DyMn$_2$O$_5$,\cite{NH04b,DH04,WR05,RE08} go directly from
the $(I,0,I)$ phase into the $(C,0,C)$ phase without the appearance
of the $(I,0,C)$ phase.  At lower temperature the 125's follow
various scenarios in which the magnetic structures may be either IC
or CM with a long period and they may or may not be ferroelectric.
For a review of the properties and Landau theory for 125's see
Ref. \onlinecite{FOCUS}.

Here we give a symmetry analysis of the allowed magnetic structures
in the $(I,0,I)$ or $(I,0,C)$ phases.  (A detailed symmetry analysis
applicable to the phase with $q_x=1/2$\cite{RC07,ABH}
indicated that this phase was described by a two dimensional (2D) irrep
and therefore could be characterized by two complex-valued order
parameters\cite{ABH} we will call $\sigma_1$ and $\sigma_2$.)  The symmetry
of the phase when $q_x \not=1/2$ is different.
The group of this wave vector contains unity $E$ and the glide
$m_{ac}$ which leaves the $b$-component of the wave vector
invariant. Thus we have two one dimensional (1D) irreps, which we
label $\Gamma_e$ and $\Gamma_o$ (``e" for even and ``o" for odd).
In particular, since
the star of the wave vector contains four vectors, ordering within
each irrep is described by four complex-valued order parameters.\cite{HAE}
The allowable wave functions are the basis functions of the irreps
which transform appropriately.  These basis functions are actually
eigenvectors of $m_{ac}$ with eigenvalues $+ \lambda^*$ (for
$\Gamma_e$) and $-\lambda^*$ (for $\Gamma_o$), where
$\lambda = \exp( -i \pi q_x)$.  Since each irrep is contained 18 times
in the original reducible representation generated by the three
spin components of the
12 magnetic sites in the unit cell (here we assume that R is magnetic),
each wave function contains 18 independent free complex-valued
parameters.  These wave functions are listed in Table \ref{IRREP},\cite{FN3}
and they are in agreement with ({\it i. e.} are a reparameterization
of) the results of the MODY program.\cite{MODY}

\begin{table}
\caption{\label{IRREP} Symmetry-adapted basis functions for
wave vector\cite{FN3}
${\bf q}_+=(q_x,0,q_z)$ which transform according to the irreps
$\Gamma_e$ and $\Gamma_o$, where $\lambda = \exp(- \pi i q_x)$.
We have not yet included the effect of inversion symmetry.}
\vspace{0.2 in}
\begin{tabular} {|| c | c | c ||}
\hline \hline
& $\psi(\Gamma_e)$ & $\psi(\Gamma_o)$ \\ \hline
${\bf S}({\bf q},1) = $ & $(s_{x1},s_{y1},s_{z1} )$ &
$( u_{x1}, u_{y1}, u_{z1}) $ \\  
${\bf S}({\bf q},2) = $ & $- \lambda (s_{x1}, - s_{y1}, s_{z1} )$
& $\lambda (u_{x1}, -u_{x2}, u_{z1}) $ \\  
${\bf S}({\bf q},3) = $ & $-\lambda^* (t_{x1}, -t_{y1}, t_{z1} )$
& $\lambda^* (v_{x1}, -v_{y1}, v_{z1}) $ \\
${\bf S}({\bf q},4) = $ & $(t_{x1},t_{y1},t_{z1} )$ &
$( v_{x1}, v_{y1}, v_{z1}) $ \\  
${\bf S}({\bf q},5) = $ & $(s_{x2},s_{y2},s_{z2} )$ &
$( u_{x2}, u_{y2}, u_{z2}) $ \\  
${\bf S}({\bf q},6) = $ & $-\lambda^* (s_{x2}, -s_{y2}, s_{z2} )$ 
& $\lambda^* (u_{x2}, -u_{y2}, u_{z2}) $ \\  
${\bf S}({\bf q},7) = $ & $-\lambda^* (t_{x2}, -t_{y2}, t_{z2} )$
& $\lambda^* (v_{x2}, - v_{y2}, v_{z2}) $ \\  
${\bf S}({\bf q},8) = $ & $(t_{x2},t_{y2},t_{z2} )$ &
$( v_{x2}, v_{y2}, v_{z2}) $ \\  
${\bf S}({\bf q},9) = $ & $(s_{x3},s_{y3},s_{z3} )$ &
$( u_{x3}, u_{y3}, u_{z3}) $ \\  
${\bf S}({\bf q},10) = $ & $-\lambda (s_{x3}, -s_{y3}, s_{z3} )$ 
& $\lambda (v_{x3}, -v_{y3}, u_{z3}) $ \\  
${\bf S}({\bf q},11) = $ & $-\lambda^* (t_{x3}, -t_{y3}, t_{z3} )$
& $\lambda^* (v_{x3}, -v_{y3}, u_{z3})
$ \\  ${\bf S}({\bf q},12) = $ & $(t_{x3},t_{y3},t_{z3} )$ &
$( v_{x3}, v_{y3}, v_{z3}) $ \\  
\hline
\end{tabular}
\end{table}

To illustrate
the transformation laws, we perform a partial check that the vectors
in Table \ref{IRREP} are indeed eigenfunctions of $m_{ac}$.  Note
that we use so-called ``unit-cell" Fourier transforms whereby\cite{FN2}
\begin{eqnarray}
{\bf S}({\bf R},n) &=& {\bf S}({\bf q},n)e^{-2 \pi i {\bf q} \cdot
{\bf R}} \ + \ {\rm c. \ c.}  \ ,
\label{TRANS1} \end{eqnarray}
where $n$ labels the sublattice and ${\bf R}$ locates the unit cell.
A transformation ${\cal O}$ takes the ``initial'' basis function into
a ``final'' basis function.  If a prime indicates ``final,"
{\it i. e.} ``after transformation," then $S'({\bf R}_f,n_f)$
denotes the spin of sublattice $n_f$ in the unit cell at
${\bf R}_f$ after transformation.  This quantity is obtained by
applying the transformation to the spin at the initial location
${\bf R}_i + \tauv_{n_i}$. Thus for transformation by $m_{ac}$ we write
\begin{eqnarray}
S'({\bf q}, n_f) &=& \xi_\alpha S_\alpha ({\bf q}, n_i)
e^{2 \pi i {\bf q} \cdot [{\bf R}_f - {\bf R}_i]} \ ,
\label{TRANS2} \end{eqnarray}
where $\xi_\alpha$ is the appropriate factor for the mirror
operation $m_{ac}$ on the components of a pseudovector:
$\xi_y = - \xi_x = - \xi_z= 1$.  We will check that the basis vector
of irrep $\Gamma_e$ is an eigenfunction of $m_{ac}$. Note that under
$m_{ac}$ when the initial sublattice index is $n_i=2n-1$, then the final
sublattice index is $n_f=2n$ and vice versa.  Thus
\begin{eqnarray}
S'_\alpha ({\bf q},1) &=& \xi_\alpha S_\alpha ({\bf q},2)
e^{2 \pi i {\bf q} \cdot [{\bf R}_f - {\bf R}_i]} \nonumber \\
&=& \xi_\alpha \lambda \xi_\alpha s_{\alpha 1} e^{2 \pi i q_x}
= \lambda^* s_{\alpha 1} = \lambda^* S_\alpha ({\bf q},1) \ , \nonumber \\
S'_\alpha ({\bf q},2) &=& \xi_\alpha S_\alpha ({\bf q},1)
e^{2 \pi i {\bf q} \cdot [{\bf R}_f - {\bf R}_i]} \nonumber \\
&=& \xi_\alpha s_{\alpha 1} = \lambda^* [\lambda \xi_\alpha s_{\alpha 1}]
= \lambda^* S_\alpha ({\bf q},2) \ , \nonumber \\
S'_\alpha ({\bf q},3) &=& \xi_\alpha S_\alpha ({\bf q},4)
e^{2 \pi i {\bf q} \cdot [{\bf R}_f - {\bf R}_i]} \nonumber \\
&=& \xi_\alpha t_{\alpha 1} e^{2 \pi i q_x} = 
\lambda^* [\lambda^* \xi_\alpha t_{\alpha 1}]
= \lambda^* S_\alpha ({\bf q},3) \ , \nonumber \\
S'_\alpha ({\bf q},4) &=& \xi_\alpha S_\alpha ({\bf q},3) 
e^{2 \pi i {\bf q} \cdot [{\bf R}_f - {\bf R}_i]} \nonumber \\
&=& \lambda^* \xi_\alpha t_{\alpha 1} = \lambda^* S_\alpha ({\bf q},4) \ ,
\nonumber \\
S'_\alpha ({\bf q},5) &=& \xi_\alpha S_\alpha ({\bf q},6)
e^{2 \pi i {\bf q} \cdot [{\bf R}_f - {\bf R}_i]} \nonumber \\
&=& \lambda^* \xi_\alpha^2 s_{\alpha 2} = \lambda^* S_\alpha ({\bf q},5) \ ,
\nonumber \\
S'_\alpha ({\bf q},6) &=& \xi_\alpha S_\alpha ({\bf q},5)
e^{2 \pi i {\bf q} \cdot [{\bf R}_f - {\bf R}_i]} \nonumber \\
&=& \xi_\alpha s_{\alpha 2} e^{2 \pi i q_x}= 
\lambda^* [\lambda^* \xi_\alpha s_{\alpha 2}]
= \lambda^* S_\alpha ({\bf q},6)   \ , \nonumber \\
S'_\alpha ({\bf q},7) &=& \xi_\alpha S_\alpha ({\bf q},8)
e^{2 \pi i {\bf q} \cdot [{\bf R}_f - {\bf R}_i]} \nonumber \\
&=& \xi_\alpha t_{\alpha 2} e^{2 \pi i q_x} 
= \lambda^* [\xi_\alpha \lambda^* t_{\alpha 2}]
= \lambda^* S_\alpha ({\bf q},7) \ , \nonumber \\
S'_\alpha ({\bf q},8) &=& \xi_\alpha S_\alpha ({\bf q},7)
e^{2 \pi i {\bf q} \cdot [{\bf R}_f - {\bf R}_i]} \nonumber \\
&=& \xi_\alpha \lambda^* \xi_\alpha t_{\alpha 2}
= \lambda^* t_{\alpha 2} = \lambda^* S_\alpha ({\bf q},8) \ , \nonumber \\
S'_\alpha ({\bf q},9) &=& \xi_\alpha S_\alpha ({\bf q},10)
e^{2 \pi i {\bf q} \cdot [{\bf R}_f - {\bf R}_i]} \nonumber \\
&=& \xi_\alpha \lambda s_{\alpha 3}e^{2 \pi i q_x} = \lambda^* 
[\xi_\alpha s_{\alpha 3}] = \lambda^* S_\alpha ({\bf q},9) \ , \nonumber \\
S'_\alpha ({\bf q},10) &=& \xi_\alpha S_\alpha ({\bf q},9)
e^{2 \pi i {\bf q} \cdot [{\bf R}_f - {\bf R}_i]} \nonumber \\
&=& \xi_\alpha s_{\alpha 3} = \lambda^* [\lambda \xi_\alpha s_{\alpha 3}]
= \lambda^* S_\alpha ({\bf q},10) \ , \nonumber \\
S'_\alpha ({\bf q},11) &=& \xi_\alpha S_\alpha ({\bf q},12)
e^{2 \pi i {\bf q} \cdot [{\bf R}_f - {\bf R}_i]} \nonumber \\
&=& \xi_\alpha t_{\alpha 3} e^{2 \pi i q_x} = \lambda^* 
[\lambda^* \xi_\alpha t_{\alpha 3}] = \lambda^* S_\alpha ({\bf q},11)
\ , \nonumber \\
S'_\alpha ({\bf q},12) &=& \xi_\alpha S_\alpha ({\bf q},11)
e^{2 \pi i {\bf q} \cdot [{\bf R}_f - {\bf R}_i]} \nonumber \\
&=& \lambda^* \xi_\alpha^2 t_{\alpha 3} = \lambda^* S_\alpha ({\bf q},12) \ .
\end{eqnarray}
Thus $\psi (\Gamma_e)$  is an eigenvector of $m_{ac}$ with eigenvalue
$\lambda^*$.  In the other irrep, the fact that $\lambda$ is everywhere 
replaced by $-\lambda$ ensures that $\psi(\Gamma_o)$ is an
eigenvector of $m_{ac}$ with eigenvalue $- \lambda^*$.

\subsection{Effect of Inversion Symmetry}

Now we modify the above results to take account of inversion
symmetry. A straightforward, if clumsy, way to do this is
to use the fact that the inverse susceptibility matrix
becomes singular at a continuous phase transition, which implies
that one of its eigenvalues passes through zero.  We wish to see
what restrictions symmetry places on the associated critical eigenvector.
We write the quadratic terms in the free energy $F_2$ in the form
\begin{eqnarray}
F_2 &=&  \frac{1}{2} \Psi^\dagger {\cal F} \Psi \ ,
\end{eqnarray}
where ${\cal F}$ is the inverse suscptibility matrix.
Instead of considering the quadratic form in the original spin
variables, we consider the quadratic form in terms of the
variables of Table \ref{IRREP}.  So the matrix ${\cal F}$ is an
18 dimensional Hermitian matrix operating on an 18-component
vector $\Psi(\Gamma)$  which we write as $({\bf s}_1, {\bf t}_1, {\bf s}_2,
{\bf t}_2, {\bf s}_3, {\bf t}_3)$, where the ${\bf s}$'s and
${\bf t}$'s are three component subvectors taken from
Table \ref{IRREP}.  Thus
\begin{eqnarray}
{\bf s}_n \equiv  (s_{xn}, s_{yn}, s_{zn}) \ .
\label{NOTE} \end{eqnarray}
Because the paramagnetic phase has symmetry under spatial inversion
${\cal I}$, we must have\cite{NVO,TMO,ABH}
\begin{eqnarray}
F_2 &=& \frac{1}{2} [{\cal I} \Psi ]^\dagger {\cal F} [ {\cal I} \Psi ] 
= \frac{1} {2} \Psi^\dagger {\cal F} \Psi \ ,
\label{INVEQ} \end{eqnarray}
for all values of the spin coordinates.  

To implement this we note that for transformation  under ${\cal I}$,
the result follows a logic similar to that leading to Eq. 
(\ref{TRANS2}), namely\cite{ABH}
\begin{eqnarray}
S'_\alpha ({\bf q}, \tau_f)^* =S_\alpha ({\bf q}, \tau_i)
e^{2 \pi i {\bf q} \cdot [ \tauv_f + \tauv_i]} \ ,
\label{EQ22} \end{eqnarray}
where again the prime indicates the value after transformation
by ${\cal I}$.  Note that inversion relates sites (1,4), (2,3),
(5,8), (6,7), (9,12), and (10,11).  Now use Eq. (\ref{EQ22}) to get
\begin{eqnarray}
s_{\alpha 1}' &=& t_{\alpha 1}^* e^{-2 \pi i(q_x+q_z)} \nonumber \\
s_{\alpha 2}' &=& t_{\alpha 2}^* e^{-2 \pi i(q_x+q_z)} \nonumber \\
s_{\alpha 3}' &=& t_{\alpha 3}^* e^{-2 \pi iq_x} 
\end{eqnarray}
and
\begin{eqnarray}
t_{\alpha 1}' &=& s_{\alpha 1}^* e^{-2 \pi i(q_x+q_z)} \nonumber \\
t_{\alpha 2}' &=& s_{\alpha 2}^* e^{-2 \pi i(q_x+q_z)} \nonumber \\
t_{\alpha 3}' &=& s_{\alpha 3}^* e^{-2 \pi iq_x} \ .
\end{eqnarray}
These simple results arise because we reparametrized with an eye
to avoiding complexity.

The eigenvalue equation for the $18 \times 18$ matrix ${\cal F}$
can be represented as
\begin{eqnarray}
\left[ \begin{array} {c c c c c c}
{\bf A} & {\bf B} & {\bf C} & {\bf D} & {\bf E} & {\bf F} \\
{\bf B}^\dagger & {\bf G} & {\bf H} & {\bf I} & {\bf J} & {\bf K} \\
{\bf C}^\dagger & {\bf H}^\dagger & {\bf L} & {\bf M} & {\bf N} & {\bf O} \\
{\bf D}^\dagger & {\bf I}^\dagger & {\bf M}^\dagger & {\bf P} & {\bf Q} 
& {\bf R} \\
{\bf E}^\dagger & {\bf J}^\dagger & {\bf N}^\dagger & {\bf Q}^\dagger &
{\bf S} &  {\bf T} \\
{\bf F}^\dagger & {\bf K}^\dagger & {\bf O}^\dagger & {\bf R}^\dagger &
{\bf T}^\dagger & {\bf U} \\
\end{array} \right] \left[ \begin{array} {c}
{\bf s}_1 \\ {\bf t}_1 \\ {\bf s}_2 \\ {\bf t}_2 \\
{\bf s}_3 \\ {\bf t}_3 \\ \end{array} \right] = \mu
\left[ \begin{array} {c}
{\bf s}_1 \\ {\bf t}_1 \\ {\bf s}_2 \\ {\bf t}_2 \\
{\bf s}_3 \\ {\bf t}_3 \\ \end{array} \right] \ ,
\end{eqnarray}
where each entry of the matrix is itself a $3 \times 3$ submatrix.
Now we identify the symmetry of this matrix imposed by inversion,
via Eq.  (\ref{INVEQ}).  We have 
\begin{eqnarray}
A_{ij} s_{i 1}^* s_{j 1} &=&
A_{ij} [{\cal I} s_{i 1}]^* [{\cal I} s_{j 1}]
= A_{ij} t_{i 1} t_{j 1}^* \nonumber \\
&=&   G_{ji } t_{j 1}^* t_{i 1 } \ ,
\end{eqnarray}
which implies that $A_{ij} = G_{ji}$, so that
${\bf G}=\tilde {\bf A}={\bf A}^*$, since ${\bf A}$ is Hermitian.
Similarly, ${\bf P} = {\bf L}^*$ and ${\bf U}={\bf S}^*$.  Consider
\begin{eqnarray}
B_{ij} s_{i1}^* t_{j1} &=& B_{ij} [{\cal I} s_{i1}]^* [{\cal I} t_{j1}] =
B_{ij} t_{i1} s_{j1}^* \nonumber \\ &=& B_{ji} s_{j1}^* t_{i1} \ ,
\end{eqnarray}
which implies that $B_{ij}=B_{ji}$.  Thus ${\bf B}^\dagger = {\bf B}^*$.
Likewise ${\bf M}^\dagger={\bf M}^*$ and ${\bf T}^\dagger = {\bf T}^*$.
Furthermore
\begin{eqnarray}
C_{ij} s_{i1}^* s_{j2} &=& C_{ij} [{\cal I} s_{i1}]^* [{\cal I}s_{j2}]
= C_{ij} t_{i1} t_{j2}^* \nonumber \\ &=& {I^\dagger}_{ji} t_{j2}^* t_{i1}
\end{eqnarray}
which implies that $I_{ij}^* = C_{ij}$.  Also
\begin{eqnarray}
E_{ij} s_{i1}^* s_{j3} &=& E_{ij} [{\cal I} s_{i1}]^* [{\cal I} s_{j3}]
\nonumber \\ &=& E_{ij} t_{i1} t_{j3}^* [ e^{2 \pi i(q_x+q_z)}]
e^{-2 \pi i q_x} \nonumber \\ &=& [K^\dagger]_{ji} t_{j3}^* t_{i1}
\end{eqnarray}
which implies that ${\bf K}^* = {\bf E} e^{2 \pi i q_z}$.  Similarly
${\bf R}^* = {\bf N} e^{2 \pi i q_z}$.  Also
\begin{eqnarray}
D_{ij} s_{i1}^* t_{j2} &=& D_{ij} [{\cal I} s_{i1}]^* [{\cal I} t_{j2}]
= D_{ij} t_{i1} s_{j2}^* \nonumber \\ &=& [H^\dagger]_{ji} s_{j2}^* t_{i1}
\end{eqnarray}
so ${\bf H}^*= {\bf D}$. Also
\begin{eqnarray}
J_{ij} t_{i1}^* s_{j3} &=& J_{ij} [{\cal I} t_{i1}]^* [{\cal I} s_{j3}]
\nonumber \\ &=&
 J_{ij} s_{i1} t_{j3}^* [e^{2 \pi i (q_x+q_z)}] e^{- 2 \pi i q_z}
\nonumber \\ &=& F^\dagger_{ji} t_{j3}^* s_{i1} \ ,
\end{eqnarray} 
which implies that ${\bf F}^* = {\bf J} e^{2 \pi i q_z}$.
Similarly ${\bf O}^* = {\bf Q} e^{2 \pi i q_z}$.

Using all these relations we see that the matrix ${\cal F}$ must be of
the form
\begin{eqnarray}
\left[ \begin{array} {c c c c c c}
 {\bf A}  & {\bf B} & {\bf C} &  {\bf D} & {\bf E} & {\bf J}^*\Lambda^*  \\
{\bf B}^* & {\bf A}^* & {\bf D}^* & {\bf C}^* & {\bf J} & {\bf E}^* \Lambda^* \\
\ \tilde {\bf C}^* & \tilde {\bf D} & {\bf L} & {\bf M}  & {\bf N}  &
{\bf Q}^* \Lambda^*  \\
\ \tilde {\bf D}^*  & \tilde {\bf C} & {\bf M}^* & {\bf L}^* & {\bf Q} &
{\bf N}^*\Lambda^* \\
\ \tilde {\bf E}^* & \tilde {\bf J}^* &\tilde {\bf N}^* &\tilde {\bf Q}^*
& {\bf S}  & {\bf T} \\
\ \tilde {\bf J}\Lambda & \tilde {\bf E} \Lambda &\tilde {\bf Q} \Lambda
& \tilde {\bf N} \Lambda & {\bf T}^*  & {\bf S}^*   \\
\end{array} \right] \ ,
\end{eqnarray}
where $\Lambda= \exp( 2 \pi i q_z)$.  Now consider this matrix
operating on a vector of the form
\begin{eqnarray}
\Psi &=& [ \rhov , \rhov^* , \psiv, \psiv^* , \chiv , \Lambda \chiv^* ] \ .
\end{eqnarray}
One can show that ${\cal F}\Psi$ is a vector of the same form as $\Psi$.
This means that any eigenvector can be taken to be of this form and the
eigenvalue equations are
\begin{eqnarray}
&& {\bf E}^\dagger \rhov + {\bf J}^\dagger \rhov^* + {\bf N}^\dagger \psiv
+ {\bf Q}^\dagger \psiv^* + {\bf S} \chiv
+ {\bf T} \Lambda \chiv^* = \lambda \chiv \nonumber \\
&& {\bf C}^\dagger \rhov + \tilde {\bf D} \rhov^* + {\bf L} \psiv
+ {\bf M} \psiv^* + {\bf N} \chiv + {\bf Q}^* \chiv^*
= \lambda \psiv \nonumber \\
&& {\bf A} \rhov + {\bf B} \rhov^* + {\bf C} \psiv + {\bf D} \psiv^* + 
{\bf E} \chiv + {\bf J}^* \chiv^* = \lambda \rhov \ .
\end{eqnarray}
(The other three equations are the complex conjugates of these.)
These give rise to 18 simultaneous equations for the real and imaginary
parts of the three-component vectors $\rhov$, $\psiv$, and $\chiv$.

The point is that the permissible form for an 18-component 
eigenvector is restricted by inversion symmetry.
The critical eigenvector is the one whose eigenvalue
first passes through zero as the temperature is lowered. As the
temperature is further lowered, we may have a small amount of
admixing of noncritical eigenvectors into the critical eigenvector
due to higher than quadratic terms in the free energy.  However, these
admixtures will only be within the same irrep unless one crosses
a phase boundary.

Since the eigenvalue problem is in a complex vector space we write
critical eigenvector as
\begin{eqnarray}
\Psi &=& e ^{i \phi} [ \rhov, \rhov^* ,
\psiv, \psi^* , \chiv , \Lambda \chiv^* ] \ ,
\end{eqnarray}
where the phase $\phi$ is arbitrary (as far as the quadratic terms
are concerned) and the other Greek letters are three component vectors.
In Tables \ref{IIRREP} and \ref{IIIRREP} we tabulate the results.  In
so doing we have introduced the complex-valued order parameters
$\sigmav (\Gamma)$, such that
\begin{eqnarray}
\sigmav (\Gamma) &=& |\sigma(\Gamma)| e^{i \phi(\Gamma)} \ .
\end{eqnarray}
To avoid overparametrizing we specify the normalization
\begin{eqnarray}
4 \sum_{\alpha=x,y,z} \sum_{n=1,2,3} | s_{\alpha n}|^2 = 1 \ .
\label{NORM} \end{eqnarray}
Including inversion symmetry we have 9 complex-valued $s$ parameters
and one complex valued order parameter $\sigma_e({\bf q}_+)$, so that
we have 19 real valued parameters (taking account of the normalization
of the $s$'s.), whereas without taking account of inversion symmetry we
would have had 36 real valued parameters to determine from a fit to 
diffraction data.

One may notice that we could have said that the 18-component eigenvector
of ${\bf u}$'s  was of the form
\begin{eqnarray}
\Phi &=& e^{i \phi}
[ \piv, -\piv^* , \tauv, -\tauv^* , \xiv , -\Lambda \xiv^* ]
\end{eqnarray}
and indeed the eigenvector is equivalent to this form because if you
multiply the previous eigenvector $\Psi$ by $i$, it will be exactly
of the form of $\Phi$.

\begin{table}
\caption{\label{IIRREP} Symmetry-adapted spin wave functions for
wave vector ${\bf q}_+ \equiv (q_x,0,q_z)$ which transform according
to the irrep $\Gamma_e$, where $\lambda = \exp(- \pi i q_x)$,
$\Lambda = \exp(2 \pi i q_z)$, and $\sigmav_e$ is the
complex-valued order parameter. We require the normalization of
Eq. (\ref{NORM}). Otherwise, all constants assume arbitrary
complex values. Here we include the effect of inversion symmetry.}
\vspace{0.2 in}
\begin{tabular} {|| c | c ||}
\hline \hline
& $\psi(\Gamma_e)$ \\ \hline
${\bf S}({\bf q},1) = $ & $\sigmav_e({\bf q}_+) (s_{x1},s_{y1},s_{z1} )$ \\  
${\bf S}({\bf q},2) = $ & $-\sigmav_e({\bf q}_+) \lambda
(s_{x1}, -s_{y1}, s_{z1} )$ \\
${\bf S}({\bf q},3) = $ & $-\sigmav_e({\bf q}_+) \lambda^*
(s_{x1}^*, -s_{y1}^*, s_{z1}^* )$ \\
${\bf S}({\bf q},4) = $ & $\sigmav_e({\bf q}_+)
(s_{x1}^*, s_{y1}^*, s_{z1}^* )$ \\
${\bf S}({\bf q},5) = $ & $\sigmav_e({\bf q}_+)  (s_{x2},s_{y2},s_{z2} )$ \\
${\bf S}({\bf q},6) = $ & $-\sigmav_e({\bf q}_+)  \lambda^*
(s_{x2}, -s_{y2}, s_{z2} )$ \\
${\bf S}({\bf q},7) = $ & $-\sigmav_e({\bf q}_+)  \lambda^*
(s_{x2}^*, -s_{y2}^*, s_{z2}^* )$ \\
${\bf S}({\bf q},8) = $ & $\sigmav_e({\bf q}_+) 
(s_{x2}^*,s_{y2}^*,s_{z2}^* )$ \\
${\bf S}({\bf q},9) = $ & $\sigmav_e({\bf q}_+)  (s_{x3},s_{y3},s_{z3} )$ \\
${\bf S}({\bf q},10) = $ & $-\sigmav_e({\bf q}_+)  \lambda
(s_{x3}, -s_{y3}, s_{z3} )$ \\
${\bf S}({\bf q},11) = $ & $-\sigmav_e({\bf q}_+)  \lambda^* \Lambda
(s_{x3}^*, -s_{y3}^*, s_{z3}^* )$\\
${\bf S}({\bf q},12) = $ & $\sigmav_e({\bf q}_+) 
\Lambda (s_{x3}^*,s_{y3}^*,s_{z3}^* )$ \\
\hline
\end{tabular}
\end{table}

\begin{table}
\caption{\label{IIIRREP} As Table \ref{IIRREP}, but for the irrep
$\Gamma_o$ and we require the normalization of
Eq. (\ref{NORM}) with $s$ replaced by $u$.}
\vspace{0.2 in}
\begin{tabular} {|| c | c ||}
\hline \hline
& $\psi(\Gamma_o)$ \\ \hline
${\bf S}({\bf q},1) = $ & $\sigmav_o({\bf q}_+) ( u_{x1}, u_{y1}, u_{z1}) $ \\  
${\bf S}({\bf q},2) = $ & $\sigmav_o({\bf q}_+) \lambda
(u_{x1}, -u_{x2}, u_{z1}) $ \\  
${\bf S}({\bf q},3) = $ & $\sigmav_o({\bf q}_+) \lambda^* (u_{x1}^*,
-u_{y1}^*, u_{z1}^* ) $ \\
${\bf S}({\bf q},4) = $ & $\sigmav_o ({\bf q}_+)
( u_{x1}^*, u_{y1}^*, u_{z1}^* ) $ \\  
${\bf S}({\bf q},5) = $ & $\sigmav_o({\bf q}_+) ( u_{x2}, u_{y2}, u_{z2}) $ \\  
${\bf S}({\bf q},6) = $ & $\sigmav_o({\bf q}_+) \lambda^* (u_{x2},
-u_{y2}, u_{z2}) $ \\  
${\bf S}({\bf q},7) = $ & $\sigmav_o({\bf q}_+) \lambda^* (u_{x2}^*,
-u_{y2}^*, u_{z2}^*) $ \\  
${\bf S}({\bf q},8) = $ & $\sigmav_o({\bf q}_+)
( u_{x2}^*, u_{y2}^*, u_{z2}^*) $ \\  
${\bf S}({\bf q},9) = $ & $\sigmav_o({\bf q}_+) ( u_{x3}, u_{y3}, u_{z3}) $ \\  
${\bf S}({\bf q},10) = $ & $\sigmav_o({\bf q}_+) \lambda
(u_{x3}, -u_{y3}, u_{z3}) $ \\  
${\bf S}({\bf q},11) = $ & $\sigmav_o({\bf q}_+) \lambda^* \Lambda (u_{x3}^*,
-u_{y3}^* , u_{z3}^*)
$ \\  ${\bf S}({\bf q},12) = $ & $\sigmav_o({\bf q}_+) \Lambda
( u_{x3}^*, u_{y3}^*, u_{z3}^* ) $ \\  
\hline
\end{tabular}
\end{table}

The comparison with Ni$_3$V$_2$O$_8$\cite{NVO,PRB,ABH} (NVO)
and TbMnO$_3$\cite{TMO,ABH} (TMO) is significant.  In the case of NVO
the magnetic Ni sites are of two types, spine and cross-tie.\cite{PRB}
All sites of the same type are related to one another
by a symmetry operation which leaves the wave vector invariant.  It
happens that the Wyckoff orbit of this set of operators generates the
entire set of spine sites and also separately the entire set of
cross-tie sites.  In that case inversion (which does not leave the wave
vector invariant) fixes all the relative phases.\cite{NVO,PRB,ABH} (The 
phases are not necessarily the same, but they are fixed.) In the case 
of TMO the Mn sites form a Wyckoff orbit of the symmetry operations that
leave the wave vector invariant, but the Tb sites break into two 
orbits.  In this case inversion fixes the relative phases within
the Mn orbit and within a single Tb orbit.  Inversion connects the two
Tb orbits.  As a result the amplitudes of the two Tb orbits are fixed
to be the same and they have phases which are the negatives of one another,
but the magnitude of this phase is arbitrary.\cite{TMO,ABH}
Here the Mn$^{3+}$, Mn$^{4+}$, and
RE sites each break up into two orbits which are interconnected
by inversion.  So it is not surprising that this situation
is like that of the Tb sites in TMO: the magnitudes of the two
related orbits, which according to MODY were unrelated, are now,
by virtue of inversion symmetry, fixed to be the same.

\section{DISCUSSION}

\subsection{Order Parameters}

It is natural to introduce order parameters because as the
temperature is reduced into the ordered phase, the critical eigenvector
is nearly temperature-independent except for a change in its
normalization, governed by the magnitude of the order parameter.
Furthermore, the phase of the complex order parameter is either
a free variable or, if it is fixed, it is only
fixed by subtle effects of higher-than-quadratic terms in the
free energy.  So the order parameter describes properly the low
energy sector of the free energy.

Note that our definition of the order parameter is such that if
one is given the spin wave function over all the sublattices it
is possible to uniquely determine both the phase and the magnitude of
the order parameter, except that it could be multiplied by $-1$.
(But that indeterminacy is inherent for this order parameter
symmetry.) To make this unique identification
from a knowledge of the wave functions, the wave functions must be
first put into the canonical form of Tables \ref{IIRREP} and \ref{IIIRREP}.
In so doing, the normalization condition has to be obeyed. Then
the prefactor will be the desired order parameter. Note that the phase
is fixed by having the first and fourth components written in terms
of complex conjugates.  This type of identification would not be
possible for a one-component complex variable.

It should be noted that the order parameter inherits the symmetry of the full
wave function.  Having the basis functions for\cite{FN3}
${\bf q}_+\equiv (q_x,0,q_z)$
we now obtain the basis functions for the other wave vectors in the
star of ${\bf q}$.  We first obtain the basis functions for 
$-{\bf q}_- = (-q_x,0,q_z)$ for irrep $\Gamma_e$.
The most general basis function
for irrep $\Gamma_e$ for this wave vector will be of the form of Table
\ref{IIRREP} with $q_x$ replaced by $-q_x$, {\it i. e.} with $\lambda$
replaced by $\lambda^*$ and, for notational convenience, $s_{\alpha,n}$
replaced by $t_{\alpha,n}$,  However, this is not the basis function we want.
We want the particular basis function which is obtained from that of
${\bf q}_+$ by a symmetry operation which takes ${\bf q}_+ = (q_x,0,q_z)$ 
into $-{\bf q}_- = (-q_x,0,q_z)$,\cite{JS06} because it is this basis
function which results from the actual interaction between spins.  In other
words, we want to relate $t_{\alpha,n}$ to $s_{\alpha,n}$.  To do this we
now study the transformation of the spin Fourier transforms.

 We first consider transformation by $2_c$ which takes ${\bf q}=(q_x,0,q_z)$
into ${\bf q}'=(-q_x,0,q_z)=-{\bf q}_-$, where here and below we use a prime to
indicate a quantity after transformation.  We have that 
\begin{eqnarray}
S'_\alpha ({\bf R}_f,1) = \rho_\alpha S_\alpha ({\bf R}_i,4)\ ,
\end{eqnarray}
where $\rho_x=\rho_y=-\rho_z=-1$. We now write this in terms of Fourier
components using Eq. (\ref{TRANS1}).   The initial position is ${\bf r}_i=
(X,Y,Z) + \tauv_4$ and the final position is ${\bf r}_f = (\overline X -1,
\overline Y -1, Z) + \tauv_1$ which gives [with $\eta= \sigma_e(-{\bf q})=
\sigma_e({\bf q}_-)^*$ and $\sigma = \sigma_e({\bf q}_+)$]
\begin{eqnarray}
&& \eta' t_{\alpha1} e^{-2 \pi i[(-q_x,0,q_z) \cdot (-X-1,-Y-1,Z)]}
\nonumber \\ &=& \rho_\alpha \sigma s_{\alpha 1}^* e^{-2 \pi i[(q_x,0,q_z)
\cdot (X,Y,Z)]} \ .
\end{eqnarray}
So with $\exp [-2 \pi i q_x]=\lambda^2$, we have
\begin{eqnarray}
\eta' \lambda^2 t_{\alpha 1} = \rho_\alpha \sigma s_{\alpha 1}^* \ .
\label{AA} \end{eqnarray}
Similarly
\begin{eqnarray}
S_\alpha' ({\bf R}_f,4) = \rho_\alpha S_\alpha ({\bf R}_i,1)
\end{eqnarray}
with ${\bf r}_i = (X,Y,Z) + \tauv_1$, ${\bf r}_f =  (\overline X -1,
\overline Y -1, Z) + \tauv_4$ which gives
\begin{eqnarray}
&& \eta' t_{\alpha1}^* e^{-2 \pi i[(-q_x,0,q_z) \cdot (-X-1,-Y-1,Z)]}
\nonumber \\ &=& \rho_\alpha \sigma s_{\alpha 1} e^{-2 \pi i[(q_x,0,q_z)
\cdot (X,Y,Z)]}
\end{eqnarray}
so that
\begin{eqnarray}
\eta' \lambda^2 t_{\alpha 1}^* = \rho_\alpha \sigma s_{\alpha 1} \ .
\label{BB} \end{eqnarray}
Similarly
\begin{eqnarray}
S_\alpha' ({\bf R}_f,5) = \rho_\alpha S_\alpha ({\bf R}_i,5)
\end{eqnarray}
with ${\bf r}_i = (X,Y,Z) + \tauv_5$, ${\bf r}_f =  (\overline X -1,
\overline Y , Z) + \tauv_5$ which is
\begin{eqnarray}
&& \eta' t_{\alpha2} e^{-2 \pi i[(-q_x,0,q_z) \cdot (-X-1,-Y-1,Z)]}
\nonumber \\ &=& \rho_\alpha \sigma s_{\alpha 2} e^{-2 \pi i[(q_x,0,q_z)
\cdot (X,Y,Z)]}
\end{eqnarray}
so that
\begin{eqnarray}
\eta' \lambda^2 t_{\alpha 2} = \rho_\alpha \sigma s_{\alpha 2} \ .
\label{CC} \end{eqnarray}
Similarly
\begin{eqnarray}
S_\alpha' ({\bf R}_f,7) = \rho_\alpha S_\alpha ({\bf R}_i,7)
\label{E33} \end{eqnarray}
with ${\bf r}_i = (X,Y,Z) + \tauv_7$, ${\bf r}_f =  (\overline X,
\overline Y -1, Z) + \tauv_7$.  In using Table \ref{IIRREP} we must
replace $\lambda$ by $\lambda^*$ to convert the table for the wave
vector ${\bf q}'$.  Thus Eq. (\ref{E33}) yields
\begin{eqnarray}
&& \lambda \xi_\alpha \eta' t_{\alpha2}^*
e^{-2 \pi i[(-q_x,0,q_z) \cdot (-X,-Y-1,Z)]}
\nonumber \\ &=& \rho_\alpha \xi_\alpha \lambda^*
\sigma s_{\alpha 2}^* e^{-2 \pi i[(q_x,0,q_z) \cdot (X,Y,Z)]}
\end{eqnarray}
so that
\begin{eqnarray}
\eta' \lambda^2 t_{\alpha 2}^* = \rho_\alpha \sigma s_{\alpha 2}^* \ .
\label{DD} \end{eqnarray}

\begin{table}
\caption{\label{TAMP} Amplitudes of the basis functions for the irrep
$\Gamma_e$ for the star of ${\bf q}$, where $\rho_\alpha = (-1,-1,1)$.
Here we give the basis functions for sublattices 1, 5, and 9.  The
remaining amplitudes are found by the appropriate  modification of Table
\ref{IIRREP} for the wave vector in question.  For the irrep $\Gamma_o$,
replace all the $s$'s by $u$'s and the remaining amplitudes are found
by the appropriate modification of Table \ref{IIIRREP} for
the wave vector in question.}
\begin{center}
\begin{tabular} {||c|c c c|| } \hline \hline
${\bf q}^+ = (q_x,0,q_z)$ & $s_{\alpha 1}$ & $s_{\alpha 2}$ & $s_{\alpha 3}$ \\
$-{\bf q}^- = (-q_x,0,q_z)$ & $\rho_\alpha s_{\alpha 1}^*$
& $\rho_\alpha s_{\alpha 2}$ & $\Lambda \rho_\alpha s_{\alpha 3}^*$ \\
${\bf q}^- = (q_x,0,-q_z)$ & $ \rho_\alpha s_{\alpha 1}$
& $\rho_\alpha s_{\alpha 2}^*$ & $\Lambda^* \rho_\alpha s_{\alpha 3}$ \\
$-{\bf q}^+ = (-q_x,0,-q_z)$ & $s_{\alpha 1}^*$ & $s_{\alpha 2}^*$
& $s_{\alpha 3}^*$ \\
\hline \hline
\end{tabular}
\end{center}
\end{table}

\begin{table*}
\caption{\label{TRANS} The first column gives the operation ${\cal O}$
and the column headed ${\bf v}_n$ gives the result of ${\cal O}{\bf v}_n$
where ${\bf v}$ is given in Eq. (\ref{E44}).  The last column gives the
eigenvalue of $d V_{\rm int}/dP_b$ in Eq. (\ref{E53})
under the operation ${\cal O}$.}

\begin{center}
\begin{tabular} {|| c || c | c | c | c | c | c | c | c || c ||} \hline \hline
${\cal O}$ &  $v_1$ & $v_2$ & $v_3$ & $v_4$ & $v_5$ & $v_6$ & $v_7$ &
$v_8$ & $dV_{\rm int} / d P_b$ \\
\hline $m_{ac}$ &
$\lambda^* v_1$ & $\lambda^* v_2$ & $-\lambda^* v_3$ & $-\lambda^* v_4$ &
$\lambda v_5$ & $\lambda v_6$ & $-\lambda v_7$ & $-\lambda v_8$ & $-1$ \\
$2_c$ &
$\lambda^2 v_6$ & $\lambda^2 v_5$ & $\lambda^2 v_8$ & $\lambda^2 v_7$ &
${\lambda^*}^2 v_2$ & ${\lambda^*}^2 v_1$ & ${\lambda^*}^2 v_4$
& ${\lambda^*}^2 v_3$ & $-1$ \\
${\cal I}$ &
$\lambda^2 \Lambda^* v_5$ & $\lambda^2 \Lambda v_6$ & $\lambda^2 \Lambda^* v_7$
& $\lambda^2 \Lambda v_8$ & ${\lambda^*}^2 \Lambda v_1$ & ${\lambda^*}^2
\Lambda^* v_2$ & ${\lambda^*}^2 \Lambda v_3$ & ${\lambda^*}^2 \Lambda^* v_4$
& $-1$ \\
$m_{bc}$ &
$\lambda v_6$ & $\lambda v_5$ & $-\lambda v_8$ & $-\lambda v_7$ &
$\lambda^* v_2$ & $\lambda^* v_1$ & $-\lambda^* v_4$ & $-\lambda^* v_3$
& $+1$ \\
$2_a$ &
$\lambda^* \Lambda^* v_2$ & $\lambda^* \Lambda v_1$ & $-\lambda^* \Lambda^*
v_4$ & $-\lambda^* \Lambda v_3$ & $\lambda \Lambda v_6$ & $\lambda \Lambda^*
v_5$ & $-\lambda \Lambda v_8$ & $-\lambda \Lambda^* v_7$ & $-1$ \\
$m_{ab}$ &
$\Lambda^* v_2$ & $\Lambda v_1$ & $\Lambda^* v_4$ & $\Lambda v_3$ &
$\Lambda v_6$ & $\Lambda^* v_5$ & $\Lambda v_8$ & $\Lambda^* v_7$ & $+1$ \\
$2_b$ &
$\lambda \Lambda^* v_5$ & $\lambda \Lambda v_6$ & $- \lambda \Lambda^* v_7$ &
$-\lambda \Lambda v_8$ & $\lambda^* \Lambda v_1$ & $\lambda^* \Lambda^* v_2$
& $-\lambda^* \Lambda v_3$ & $-\lambda^* \Lambda^* v_4$ & $+1$ \\
\hline \hline
\end{tabular}
\end{center}
\end{table*}

Similarly
\begin{eqnarray}
S_\alpha' ({\bf R}_f,9) = \rho_\alpha S_\alpha ({\bf R}_i,12)
\label{E36} \end{eqnarray}
with ${\bf r}_i = (X,Y,Z) + \tauv_{12}$, ${\bf r}_f =  (\overline X -1,
\overline Y -1, Z) + \tauv_9$, which gives
\begin{eqnarray}
&& \eta' t_{\alpha 3} e^{-2 \pi i[(-q_x,0,q_z) \cdot (-X-1,-Y-1,Z)]}
\nonumber \\ &=& \rho_\alpha \sigma \Lambda s_{\alpha 3}^*
e^{-2 \pi i[(q_x,0,q_z) \cdot (X,Y,Z)]}
\end{eqnarray}
so that
\begin{eqnarray}
\eta' \lambda^2 t_{\alpha 3} = \rho_\alpha \Lambda \sigma s_{\alpha 3}^* \ .
\label{EE} \end{eqnarray}
Similarly
\begin{eqnarray}
S_\alpha' ({\bf R}_f,12) = \rho_\alpha S_\alpha ({\bf R}_i,9)
\label{E39} \end{eqnarray}
with ${\bf r}_i = (X,Y,Z) + \tauv_{9}$, ${\bf r}_f =  (\overline X -1,
\overline Y -1, Z) + \tauv_{12}$, which gives
\begin{eqnarray}
&& \Lambda \eta' t_{\alpha 3}^* e^{-2 \pi i[(-q_x,0,q_z) \cdot (-X-1,-Y-1,Z)]}
\nonumber \\ &=& \rho_\alpha \sigma s_{\alpha 3}
e^{-2 \pi i[(q_x,0,q_z) \cdot (X,Y,Z)]}
\end{eqnarray}
so that
\begin{eqnarray}
\eta' \lambda^2 \Lambda t_{\alpha 3}^* = 
\rho_\alpha \Lambda \sigma s_{\alpha 3} \ .
\label{FF} \end{eqnarray}
Equations (\ref{AA}), (\ref{BB}), (\ref{CC}),
(\ref{DD}), (\ref{EE}), and (\ref{FF}) yield
\begin{eqnarray}
t_{\alpha 1} = \rho_\alpha s_{\alpha 1}^* \ , \ \ \
t_{\alpha 2} = \rho_\alpha s_{\alpha 2} \ , \ \ \
t_{\alpha 3} = \Lambda \rho_\alpha s_{\alpha 3}^* \ ,
\label{E42} \end{eqnarray}
and
\begin{eqnarray}
\eta' &=& {\lambda^*}^2 \sigma \ .
\label{E43} \end{eqnarray}
There is an equivalent solution in which all the transformed quantities
are multiplied by $-1$.  This ambiguity is unavoidable because it is
inherent in the symmetry of the order parameter.  Using Eq. (\ref{E42})
and the fact that the basis functions for $-{\bf q}$ are the complex
conjugates of those for ${\bf q}$ we obtain the results of Table \ref{TAMP}.
The relations for the basis functions of irrep $\Gamma_o$ are the
same as for $\Gamma_e$, so Table \ref{TAMP} also applies for $\Gamma_o$.

We now obtain the transformation properties of the order parameter
under all the symmetry operations of the space group (except translations).
For this discussion it is convenient to introduce an
order parameter vector ${\bf v}$ whose components are the various order
parameters:
\begin{eqnarray}
v_1 &=& \sigma_e({\bf q}_+) \ , \ \ \
v_2 = \sigma_e({\bf q}_-) \ , \ \ \
v_3 = \sigma_o({\bf q}_+) \ , \nonumber \\
v_4 &=& \sigma_o({\bf q}_-) \ , \ \ \
v_5 = \sigma_e(-{\bf q}_+) \ , \ \ \
v_6 = \sigma_e(-{\bf q}_-) \ , \nonumber \\
v_7 &=& \sigma_o(-{\bf q}_+) \ , \ \ \
v_8 = \sigma_o(-{\bf q}_-) \ .
\label{E44} \end{eqnarray}
The transformation properties of the vector ${\bf v}$ are given in 
Table \ref{TRANS}, whose construction we now discuss.  The row
of $m_{ac}$ is obtained by using the fact that the basis vector
of irrep $\Gamma_e$ for wave vector ${\bf q}=(q_x,0,q_z)$ is an 
eigenvector of $m_{ac}$ with eigenvalue $\lambda^*$.  The eigenvalue for
irreps $\Gamma_e$ and $\Gamma_o$ have opposite signs, and changing
the sign of the wave vector leads to complex conjugation of the
eigenvalue. 
 
We consider next the effect of $2_c$ on the order parameters.  In Eq. 
(\ref{E43}) we found that under $2_c$ the new value of $v_6$
is ${\lambda^*}^2 v_1$.  Since the prefactor ${\lambda^*}^2$ does not
depend on $q_z$ and it was obtained without specifying the irrep,
we see that the prefactors in the last four columns of the second
row are the same.  The prefactors of the first four entries of this
row are obtained from the last four entries by complex conjugation.

Next we consider the effect of inversion on the order parameters.  This
discussion is simplified by having in hand the results of Table
\ref{TAMP}.  Note that ${\cal I}$ does not change the orientation of
the spin, because spin is a pseudovector.  So under ${\cal I}$ we have
\begin{eqnarray}
S'_\alpha ({\bf R}_f, 1) &=& S_\alpha ({\bf R}_i, 4) \ ,
\end{eqnarray}
where ${\bf r}_i = (X,Y,Z)+ \tauv_4$ and ${\bf r}_f = (\overline X -1,
\overline Y -1, \overline Z -1) + \tauv_1$, which gives 
[with $\eta=\sigma_e(-{\bf q})$ and $\sigma= \sigma_e({\bf q})$]
\begin{eqnarray}
&& \eta' s'_{\alpha 1} e^{-2 \pi i[(-q_x,0,-q_z) \cdot (-X-1,-Y-1,-Z-1)]}
\nonumber \\
&& + {\eta'}^* s^{\prime*}_{\alpha 1} e^{2 \pi i[(-q_x,0,-q_z) \cdot
(-X-1,-Y-1,-Z-1)]} \nonumber \\
&& = \sigma s^*_{\alpha 1} e^{-2 \pi i[(q_x,0,q_z) \cdot(X,Y,Z)]}
\nonumber \\ 
&& + \sigma^* s_{\alpha1} e^{2 \pi i[(q_x,0,q_z) \cdot (X,Y,Z)]} \ .
\end{eqnarray}
This has to be an equality for all integer $X$, $Y$, and $Z$.  Also
$s'_{\alpha 1} = s_{\alpha 1}^*$ (from Table \ref{TAMP}), so we find that
\begin{eqnarray}
\eta' \lambda^2 \Lambda^* &=& \sigma \ .
\end{eqnarray}
Thus ${\lambda^*}^2 \Lambda v_1$ is the entry under $v_5$ in
the third row. Having this result, one can construct the other entries
in this row by noting the dependence on $q_x$ and $q_z$.
 
The other rows of Table \ref{TRANS} are found by using the
multiplicative properties
\begin{eqnarray}
m_{ab} &=& 2_c {\cal I} \ , \hspace{1 in} 2_a = m_{ac} m_{ab} \ , 
\nonumber \\
m_{bc} &=& 2_a {\cal I} \ , \hspace{1 in} 2_b = m_{ac} {\cal I} \  .
\end{eqnarray}

\section{Magnetoelectric Interaction}

Now we discuss the form of the ME coupling in the phases with
$q_x \not=1/2$, {\it i. e.} in the $(I,0,I)$ and $(I,0,C)$ phases.
In the first subsection we will discuss the trilinear ME interaction
which involves the lowest number (two) of magnetic order parameters.
In succeeding subsections we will discuss higher order ME interactions
which involve a product of four magnetic order parameters.  These
higher order terms yield components of the spontaneous
polarization which are allowed by symmetry but are not present
in the trilinear interaction.  However, these higher-order terms
are probably small for two reasons.  Firstly, in the IC phases
which occur at high temperatures near the paramagnetic phase, the
order parameters are small.  Secondly, most microscopic models
of the ME interaction\cite{MICRO1,MICRO2,MICRO3,MICRO4,LC06} treat
(within lowest order perturbation theory)
a trilinear Hamiltonian involving two spin variables and one
displacement variable.  However, to obtain these higher order
phenomenological interactions probably involve processes of higher
order in some small parameter such as $t/U$ or $\lambda/U$,
where $t$ is a hopping matrix element, $\lambda$ is the spin-orbit
constant, and $U$ is a Coulomb interaction. 

\subsection{Trilinear ME coupling}

Initially we will consider the lowest
order (trilinear) ME coupling. We start by considering the case when
only the wave vectors $\pm{\bf q}_+ \equiv \pm (1/2-\delta , 0,1/4 + \epsilon )$,
where $\epsilon$ may or may not be zero, are involved.
The interaction of lowest order in the magnetic order parameters
which conserves wave vector and is time-reversal invariant
is of the form\cite{NVO,TMO,ABH}
\begin{eqnarray}
V_{\rm int} &=& \sum_{\gamma a,b} c_{\gamma ab} \sigma_a ({\bf q}_+)
\sigma_b(-{\bf q}_+) P_\gamma \ ,
\end{eqnarray}
where $a$ and $b$ assume the values ``e" and ``o", ${\bf P}$ is the
spontaneous electric polarization and $\gamma$ labels the 
component.\cite{QFN}
Using Table \ref{TRANS}, one sees that terms in $V_{\rm int}$ with
$a=b$ are not allowed by inversion invariance.
If one has only a single irrep present, then one
can always redefine the location of the origin so 
as to have inversion symmetry with respect to that new origin
and hence such a phase can not exhibit magnetically
induced ferroelectricity.  If both irreps are present, then we write
\begin{eqnarray}
V_{\rm int} &=& \sum_\gamma  
[c_\gamma \sigma_e ({\bf q}_+) \sigma_o(-{\bf q}_+)
+ c_\gamma^* \sigma_e (-{\bf q}_+) \sigma_o({\bf q}_+)] P_\gamma \ ,
\nonumber \\
\end{eqnarray}
and inversion invariance forces $c_\gamma$ to be pure imaginary:
$c_\gamma = i r_\gamma$, where $r_\gamma$ is real.  Then
\begin{eqnarray}
V_{\rm int} &=& i \sum_\gamma  r_\gamma
[\sigma_e ({\bf q}_+) \sigma_o({\bf q}_+)^*
- \sigma_e ({\bf q}_+)^* \sigma_o({\bf q}_+)] P_\gamma \ .
\nonumber \\
\end{eqnarray}
From Table \ref{TRANS} one sees that the square bracket in this equation
changes sign under $m_{ac}$, so $P_\gamma$ must also change sign under
$m_{ac}$ in order for $V_{\rm int}$ to be invariant under $m_{ac}$.
Thus $c_\gamma$ can be nonzero only for $\gamma=b$, as is observed.
If we set $\sigmav_\Gamma({\bf q}_+)=
|\sigma_\Gamma({\bf q}_+)| \exp(i \phi_\Gamma)$, then we have the result
\begin{eqnarray}
V_{\rm int} &=& 2r \sin(\phi_o-\phi_e) P_b
|\sigma_e ({\bf q}_+) \sigma_o({\bf q}_+)| \ .
\end{eqnarray}

However, this is not the whole story because we must include the terms
involving the other wave vectors in the star of ${\bf q}$. (Indeed it is
possible that in the highest temperature paraeletric
IC phase there is a simultaneous condensation of the order order parameters
of both wave vectors ${\bf q}_\pm$.\cite{HAE})  Since we have already
incorporated the effect of ${\cal I}$ and $m_{ac}$, it only remains
to use $2_c$ to obtain the other terms which make up the invariant
interaction.  To do that we use the results given in Table \ref{TRANS}
which give $2_c \sigma_n({\bf q}_+) = \lambda^2 \sigma_n({\bf q}_-)^*$,
for $n=o$ or $e$, and, of course, $2_c P_b = - P_b$. Thereby we obtain
the complete result for $V_{\rm int}$:
\begin{eqnarray}
V_{\rm int} &=& ir \sum_{\eta = \pm}
[\sigmav_e ({\bf q}_\eta) \sigmav_o({\bf q}_\eta)^* 
% \nonumber \\ && \ \ \
- \sigmav_e ({\bf q}_\eta)^* \sigmav_o({\bf q}_\eta)] P_b \ . 
\nonumber \\
\label{E53} \end{eqnarray}
At this order one needs the simultaneous presence of both the e and o
irreps to have ferroelectricity.  (However, below we find that a
polarization along ${\bf c}$ can be induced by {\it Umklapp} ME
interactions by a single irrep. But this scenario is unlikely.\cite{HAE})
Note that from this interaction the spontaneous polarization ${\bf P}$ is
aligned along the ${\bf b}$ axis irrespective of which wave vector condenses.
However, the {\it sign} of ${\bf P}$ depends on how the signs of the order
parameters are chosen ({\it i. e.} how symmetry is broken) when $\sigma_o$
and/or $\sigma_e$ order.  Furthermore, within the trilinear ME interaction, 
even if two irreps are present, if they are in phase [{\it i. e.} if
$(\phi_o-\phi_e)/\pi$ is an integer], then a spontaneous polarization does not
arise.\cite{NVO,ABH} When cooling from the paramagnetic phase into the
$(I,0,I)$ phase, one expects only a {\it single} irrep.\cite{FN} Upon further
cooling, systems that follow the scenario of Fig. 1a condense a second irrep
and thereby\cite{HAE} induce ferroelectricity.  When we have both irreps of
the wave vector present, their relative phase
$[\phi(\Gamma_e) - \phi(\Gamma_o)]/\pi$ is usually fixed by fourth order
terms in the magnetic free energy to be nonintegral,\cite{ABH,HAE}
in which case no choice of origin will simultaneously make both irreps
inversion invariant.  This situation is reminiscent of TMO\cite{TMO} or
NVO\cite{NVO} and was previously noted in connection with second harmonic
generation.\cite{DF98} Finally, from Eq. (\ref{E53}) one sees that even
when two irreps are present, if the order parameters of the two wave
vectors ${\bf q}_+$ and ${\bf q}_-$ have the same magnitude, the spontaneous
polarization could vanish.  (This probably corresponds to the spirals
of the two wave vectors having opposite helicity.\cite{MOSTOVOY})

\subsection{Higher Order ME Coupling}

Sergienko {\it et al.}\cite{IS06} have pointed out the existence of
higher order terms in the ME coupling, in particular terms
quartic in the order parameters.  As they indicate, these terms have
the potential to induce a spontaneous polarization in direction(s)
different from those of the trilinear ME coupling.  For the
so-called 113 compounds (such as HoMnO$_3$, which they consider), these
terms usually do not come into play in view of the anisotropy of
the terms in the purely magnetic free energy which are quartic in the
order parameters. (See citation 28 of Ref. \onlinecite{HAE}.)
Here the situation is different: the quartic
order-parameter anisotropy is much more complicated for the 125's,
so that these higher order ME terms may come into play, although,
as mentioned, their effect may be small.  We start by first considering
terms which strictly conserve wave vector. Later, we will
investigate the corresponding {\it Umklapp} terms which only conserve
wave vector to within a nonzero reciprocal lattice vector.

\begin{table}
\begin{center}
\caption {\label{POINT} Character table for the point group
for the 125's.  $\Gamma_\alpha$, where $\alpha=x,y,z$ are vector
irreps.  The next-to-last row gives the characters of the
34-dimensional reducible representation $\Gamma$ and the last row
gives those of the 18-dimensional reducible representation, $\Gamma_U$.}

\vspace{0.2 in}
\begin{tabular} {|| c || c c c c c c c c||} \hline \hline
& $E$ & $m_{bc}$ & $m_{ac}$ & $m_{ab}$ & ${\cal I}$
& $2_a$ & $2_b$& $2_c$\\
\hline
$\Gamma_1$     & $1$ & $1$ & $1$ & $1$ & $1$ & $1$ & $1$ & $1$ \\
$\Gamma_x$     & $1$ & $-1$ & $1$ & $1$ & $-1$ & $1$ & $-1$ & $-1$ \\
$\Gamma_y$     & $1$ & $1$ & $-1$ & $1$ & $-1$ & $-1$ & $1$ & $-1$ \\
$\Gamma_z$     & $1$ & $1$ & $1$ & $-1$ & $-1$ & $-1$ & $-1$ & $1$ \\
$\Gamma_{yz}$  & $1$ & $1$ & $-1$ & $-1$ & $1$ & $1$ & $-1$ & $-1$ \\
$\Gamma_{xz}$  & $1$ & $-1$ & $1$ & $-1$ & $1$ & $-1$ & $1$ & $-1$ \\
$\Gamma_{xy}$  & $1$ & $-1$ & $-1$ & $1$ & $1$ & $-1$ & $-1$ & $1$ \\
$\Gamma_{xyz}$ & $1$ & $-1$ & $-1$ & $-1$ & $-1$ & $1$ & $1$ & $1$ \\ \hline 
$\Gamma$ & 34 & 4 & 2 & 4 & 10 & 4 & 10 & 4 \\
$\Gamma_U$ & 18 & 6 & 2 & 0 & 0 & 0 & 0 & 6 \\ \hline \hline
\end{tabular}
\end{center}
\end{table}

To construct this ME interaction we need to construct quartic
terms in the order parameters which transform like a vector.
To avoid complications, it is simplest to use the following approach
suggested by Mukamel.\cite{MK70}  The idea is to first find
the number of such vector representations by using the character
tables to determine how many times each vector irrep is contained in
the reducible representation formed by the basis functions of {\it all}
fourth order terms. The 34 fourth order terms are the nine distinct
terms of the form
\begin{eqnarray}
\sigma_k({\bf q}_+) \sigma_l({\bf q}_+) \sigma_m({\bf q}_+)^*
\sigma_n({\bf q}_+)^* \ , 
\end{eqnarray}
the nine distinct terms of the form
\begin{eqnarray}
\sigma_k({\bf q}_-) \sigma_l({\bf q}_-) \sigma_m({\bf q}_-)^*
\sigma_n({\bf q}_-)^* \ , 
\end{eqnarray}
and the 16 terms of the form
\begin{eqnarray}
\sigma_k({\bf q}_+) \sigma_l({\bf q}_-) \sigma_m({\bf q}_+)^*
\sigma_n({\bf q}_-)^* \  ,
\end{eqnarray}
where $k$, $l$, $m$, and $n$ assume the values o and e.
The character table for the irreps of the point group of Pbam and
that for the representation $\Gamma$ generated by the quartic terms
are given in Table \ref{POINT}.  The characters of the representation
$\Gamma$ for each operator are obtained by taking the trace of the
operator in the 34 dimensional vector space under consideration.

Then, we find the number of times $n(\Gamma_\alpha)$
that $\Gamma_\alpha$ is contained in $\Gamma$ is given by
the scalar products of the character vectors given in Table
\ref{POINT} as\cite{EPW}
\begin{eqnarray}
n(\Gamma_x) &=& (34 - 4 + 2 + 4 -10 + 4 -10 - 4)/8= 2 \ , \nonumber \\
n(\Gamma_y) &=& (34 + 4 - 2 + 4 -10 - 4 +10 - 4)/8= 4 \ , \nonumber \\
n(\Gamma_z) &=& (34 + 4 + 2 - 4 -10 - 4 -10 + 4)/8= 2 \ . \nonumber \\
\end{eqnarray}

We find the two $x$-like functions to be
\begin{eqnarray}
\phi_{x,1} &=& v_3^2 v_5^2 + v_4^2 v_6^2 - v_2^2 v_8^2 - v_1^2 v_7^2 
\nonumber \\ &=& [\sigma_o({\bf q}_+) \sigma_e({\bf q}_+)^*]^2
+ [\sigma_o({\bf q}_-) \sigma_e({\bf q}_-)^*]^2
\nonumber \\ && \ \
- [\sigma_e({\bf q}_-) \sigma_o({\bf q}_-)^*]^2
- [\sigma_e({\bf q}_+) \sigma_o({\bf q}_+)^*]^2 \ , \nonumber \\
\phi_{x,2} &=& v_3 v_4 v_5 v_6 - v_1 v_2 v_7 v_8
\nonumber \\ &=& \sigma_o({\bf q}_+) \sigma_o({\bf q}_-)
\sigma_e({\bf q}_+)^* \sigma_e({\bf q}_-)^*
\nonumber \\ && \ \
- \sigma_e({\bf q}_+) \sigma_e({\bf q}_-)
\sigma_o({\bf q}_+)^* \sigma_o({\bf q}_-)^* \ .
\end{eqnarray}
The above are easy to check, at least apart from the complex phase
factors which always combine to give unity.  To be invariant under $m_{ac}$
we must have an even number of ``o"'s and an even number of ``e"'s.
Note that to be odd under ${\cal I}$, the form must be odd under
complex conjugation.  To be even under $m_{ab}$ the form must be even under
interchange of ${\bf q}_+$ and ${\bf q}_-$.

We find the four $y$-like functions to be
\begin{eqnarray}
\phi_{y,1} &=& v_1 v_3 v_5^2 + v_2 v_4 v_6^2 - v_2^2 v_6 v_8 
- v_1^2 v_5 v_7 \nonumber \\ &=& |\sigma_e({\bf q}_+)|^2
[\sigma_o({\bf q}_+) \sigma_e({\bf q}_+)^*
-\sigma_o({\bf q}_+)^* \sigma_e({\bf q}_+)]
\nonumber \\ && \ \ + |\sigma_e({\bf q}_-)|^2
[\sigma_o({\bf q}_-) \sigma_e({\bf q}_-)^*
-\sigma_o({\bf q}_-)^* \sigma_e({\bf q}_-)] \nonumber \\
\phi_{y,2} &=& v_3^2 v_5 v_7 + v_4^2 v_6 v_8 - v_2 v_4 v_8^2 
- v_1 v_3 v_7^2 \nonumber \\ &=& |\sigma_o({\bf q}_+)|^2
[\sigma_o({\bf q}_+) \sigma_e({\bf q}_+)^*
-\sigma_o({\bf q}_+)^* \sigma_e({\bf q}_+)]
\nonumber \\ && \ \ + |\sigma_e({\bf q}_-)|^2
[\sigma_o({\bf q}_-) \sigma_e({\bf q}_-)^*
-\sigma_o({\bf q}_-)^* \sigma_e({\bf q}_-)] \nonumber \\
\phi_{y,3} &=& v_1 v_5 [v_4 v_6-v_8 v_2] + v_2 v_6[v_3 v_5-v_1 v_7]
\nonumber \\ &=& |\sigma_e ({\bf q}_+)|^2 
[ \sigma_o({\bf q}_-) \sigma_e({\bf q}_-)^*
- \sigma_o({\bf q}_-)^* \sigma_e({\bf q}_-)] \nonumber \\
&& \ \ + |\sigma_e ({\bf q}_-)|^2 
[ \sigma_o({\bf q}_+) \sigma_e({\bf q}_+)^*
- \sigma_o({\bf q}_+)^* \sigma_e({\bf q}_+)] \nonumber \\
\phi_{y,4} &=& v_4 v_8[v_1 v_7-v_3 v_5] + v_3 v_7[v_2 v_8-v_4 v_6]
\nonumber \\ &=& |\sigma_o({\bf q}_-)|^2
[ \sigma_e ({\bf q}_+) \sigma_o({\bf q}_+)^*
- \sigma_o ({\bf q}_+) \sigma_e({\bf q}_+)^*] \nonumber \\
&& \ \ \ + |\sigma_o({\bf q}_+)|^2 
[ \sigma_e ({\bf q}_-) \sigma_o({\bf q}_-)^*
- \sigma_o ({\bf q}_-) \sigma_e({\bf q}_-)^*] \ . \nonumber \\
\end{eqnarray}
These can be checked similarly. To be odd under $m_{ac}$ the ``e"'s and
the ``o"'s must both appear an odd number of times.

We find the two $z$-like functions to be
\begin{eqnarray}
\phi_{z,1} &=& v_3^2 v_5^2 - v_4^2 v_6^2 + v_2^2 v_8^2 - v_1^2 v_7^2 
\nonumber \\ &=& [\sigma_o({\bf q}_+) \sigma_e({\bf q}_+)^*]^2
- [\sigma_o({\bf q}_-) \sigma_e({\bf q}_-)^*]^2
\nonumber \\ && \ \ + [\sigma_e({\bf q}_-) \sigma_o({\bf q}_-)^*]^2
- [\sigma_e({\bf q}_+) \sigma_o({\bf q}_+)^*]^2 \ , \nonumber \\
\phi_{z,2} &=& v_2 v_3 v_5 v_8 - v_1 v_4 v_6 v_7
\nonumber \\ &=& \sigma_e({\bf q}_-) \sigma_o({\bf q}_+)
\sigma_e({\bf q}_+)^* \sigma_o({\bf q}_-)^*
\nonumber \\ && \ \ - \sigma_e({\bf q}_-)^* \sigma_o({\bf q}_+)^*
\sigma_e({\bf q}_+) \sigma_o({\bf q}_-) \ .
\end{eqnarray}
These can be checked similarly. To be odd under $m_{ab}$ the form must
be odd under interchange of ${\bf q}_+$ and ${\bf q}_-$.

The ME interaction of order $\sigma^4$ is written as
\begin{eqnarray}
V_{\rm ME}^{(4)} &=& \sum_{n, \gamma} c_{n,\gamma} \phi_{\gamma ,n}
P_\gamma \ ,
\end{eqnarray}
where the $c_{n,\gamma}$ are unknown coefficients.  
Now we discuss how $V_{\rm ME}^{(4)}$ affects the ME
phase diagrams.  First of all, if there is only a single irrep, either
an ``e" or an ``o", then this interaction vanishes.  So in the 
$(I,0,I)$ phase, which has only a single irrep,\cite{FN}
we still have no spontaneous polarization.
As mentioned in the introduction to this section, this
higher order ME interaction may be small and difficult to observe.

\subsection{{\it Umklapp} ME Interactions}

Now we consider {\it Umklapp} terms relevant to the phase in which 
$q_z=1/4$ but $q_x \not= 1/2$.  Here the reducible representation
$\Gamma_U$ is generated by the nine terms of the form
\begin{eqnarray}
\sigma_k({\bf q}_+) \sigma_l({\bf q}_+) \sigma_m({\bf q}_-)^*
\sigma_n({\bf q}_-)^* \delta_{4q_z,1} \ , 
\end{eqnarray}
and the nine terms of the form
\begin{eqnarray}
\sigma_k({\bf q}_-) \sigma_l({\bf q}_-) \sigma_m({\bf q}_+)^*
\sigma_n({\bf q}_+)^* \delta_{4q_z,1} \  .
\end{eqnarray}
The characters for $\Gamma_U$ are given in Table \ref{POINT}.
Then, we find the number of times $n(\Gamma_\alpha)$
that $\Gamma_\alpha$ is contained in $\Gamma_U$ to be\cite{EPW}
\begin{eqnarray}
n(\Gamma_x) &=& (18 - 6 + 2 + 0 -0 + 0 -0 - 6)/8= 1 \ , \nonumber \\
n(\Gamma_y) &=& (18 + 6 - 2 + 0 -0 - 0 +0 - 6)/8= 2 \ , \nonumber \\
n(\Gamma_z) &=& (18 + 6 + 2 - 0 -0 - 0 -0 + 6)/8= 4 \ . \nonumber \\
\end{eqnarray}
We find the $x$-like function to be
\begin{eqnarray}
\psi_{x,1} &=& v_3^2 v_6^2 + v_4^2 v_5^2 - v_2^2 v_7^2 - v_1^2 v_8^2 
\nonumber \\ &=& [\sigma_o({\bf q}_+) \sigma_e({\bf q}_+)^*]^2
+ [\sigma_o({\bf q}_-) \sigma_e({\bf q}_-)^*]^2
\nonumber \\ && \ \
- [\sigma_e({\bf q}_-) \sigma_o({\bf q}_-)^*]^2
- [\sigma_e({\bf q}_+) \sigma_o({\bf q}_+)^*]^2 \ , \nonumber \\
\end{eqnarray}
the two $y$-like functions to be
\begin{eqnarray}
\psi_{y,1} &=& v_1 v_3 v_6^2 + v_2 v_4 v_5^2 - v_2^2 v_5 v_7
- v_1^2 v_6 v_8 \nonumber \\ &=& \sigma_e({\bf q}_+)\sigma_e({\bf q}_-)^*
[\sigma_o({\bf q}_+) \sigma_e({\bf q}_-)^*
\nonumber \\ && -\sigma_e({\bf q}_+) \sigma_o({\bf q}_-)^*]
+ \sigma_e({\bf q}_-)\sigma_e({\bf q}_+)^*
\nonumber \\ && \times [\sigma_o({\bf q}_-) \sigma_e({\bf q}_+)^*
-\sigma_e({\bf q}_-) \sigma_o({\bf q}_+)^*] \nonumber \\
\psi_{y,2} &=& v_3^2 v_6 v_8 + v_4^2 v_5 v_7 - v_2 v_4 v_7^2 
- v_1 v_3 v_8^2 \nonumber \\ &=& \sigma_o({\bf q}_+)\sigma_o({\bf q}_-)^*
[\sigma_o({\bf q}_+) \sigma_e({\bf q}_-)^*
\nonumber \\ && -\sigma_o({\bf q}_-)^* \sigma_e({\bf q}_+)]
+ \sigma_o({\bf q}_-) \sigma_o({\bf q}_+)^*
\nonumber \\ && \times [\sigma_o({\bf q}_-) \sigma_e({\bf q}_+)^*
-\sigma_o({\bf q}_+)^* \sigma_e({\bf q}_-)] \ , \nonumber \\
\end{eqnarray}
and the four $z$-like functions to be
\begin{eqnarray}
\psi_{z,1} &=& v_1^2 v_6^2 - v_2^2 v_5^2
\nonumber \\ &=& [\sigma_e({\bf q}_+) \sigma_e({\bf q}_-)^*]^2
- [\sigma_e({\bf q}_-) \sigma_e({\bf q}_+)^*]^2
\nonumber \\
\psi_{z,2} &=& v_3^2 v_8^2 - v_4^2 v_7^2
\nonumber \\ &=& [\sigma_o({\bf q}_+) \sigma_o({\bf q}_-)^*]^2
- [\sigma_o({\bf q}_-) \sigma_o({\bf q}_+)^*]^2
\nonumber \\
\psi_{z,3} &=& v_1 v_3 v_6 v_8 - v_2 v_4 v_5 v_7 \nonumber \\
&=& \sigma_e({\bf q}_+) \sigma_o({\bf q}_+) 
\sigma_e({\bf q}_-)^* \sigma_o({\bf q}_-)^* \nonumber \\ && \ \
- \sigma_e({\bf q}_-) \sigma_o({\bf q}_-) 
\sigma_e({\bf q}_+)^* \sigma_o({\bf q}_+)^* \nonumber \\ && \ \
\nonumber \\
\psi_{z,4} &=& v_3^2 v_6^2 - v_2^2 v_7^2 - v_1^2 v_8^2 + v_4^2 v_5^2
\nonumber \\ &=&
[\sigma_o({\bf q}_+) \sigma_e({\bf q}_-)^*]^2
- [\sigma_e({\bf q}_-) \sigma_o({\bf q}_+)^*]^2
\nonumber \\ &+&
 [\sigma_e({\bf q}_+) \sigma_o({\bf q}_-)^*]^2
- [\sigma_o({\bf q}_-) \sigma_e({\bf q}_+)^*]^2 \ .
\end{eqnarray}
The transformation properties of the $\psi_{\alpha,n}$ can be checked
just as we did for the $\phi_{\alpha,n}$.
The {\it Umklapp} ME interaction of order $\sigma^4$ is written as
\begin{eqnarray}
V_{\rm ME,U}^{(4)} &=& \delta_{4q_z,1} 
\sum_{n, \gamma} c'_{n,\gamma} \psi_{\gamma ,n} P_\gamma \ ,
\end{eqnarray}
where the $c'_{n,\gamma}$ are unknown coefficients.

Clearly this interaction is only operative when $q_z$ is locked
to the CM value $q_z=1/4$.  This is therefore a
generalization of the term introduced by Betouras {\it et al.},\cite{BET}
but here we give the first analysis of the symmetry of this interaction.
It is interesting to note that this interaction can induce a spontaneous
polarization along the $z$-axis {\it even when only a single irrep
is present}.  (Inspection of $\psi_{z,1}$ and $\psi_{z,2}$ indicates
that this requires simultaneous condensation of order at
wave vectors ${\bf q}_\pm$.)  However, as mentioned in the introduction
to this section,
this higher order ME interactions may be small and difficult to observe.

\section{Compatibility Relations}

\begin{figure}[ht]
\begin{center}
\includegraphics[width=7.0 cm]{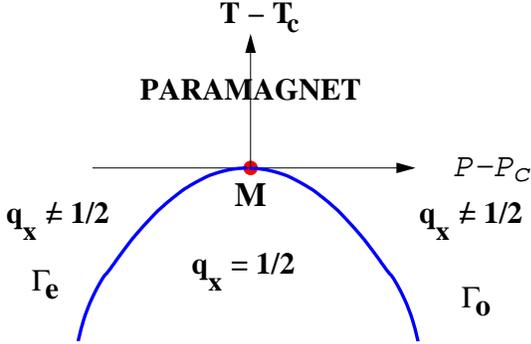}
\caption{\label{INV1} (Color online)
Phase diagram (simplified from Ref. \onlinecite{HAE}) as a function of
${\cal P}$ and $T$ for fixed $q_z$ near the multicritical point.  The phase with
$q_x=1/2$ exists within a parabolic ``tongue" whose apex is the multicritical
point M, where the $\Gamma_e$ and $\Gamma_o$ irreps interchange stability.
The compatibility relations we obtain apply in the vicinity of the
multicritical point M.}
\end{center}
\end{figure}

A first step to constructing a generic phase diagram for the 
125's\cite{HAE} is to understand how the wave functions behave 
near the phase transition between the phase with $q_x \not= 1/2$
and that for which $q_x=1/2$.  In Fig. \ref{INV1} we show
a simplified version of this phase diagram for fixed $q_z$.
(However, to compare with experiment, the diagram for fixed
$q_x$ is more relevant.\cite{HAE}) To avoid confusion we introduce a
control parameter ${\cal P}$ such that when ${\cal P}={\cal P}_c$ 
the wave vector which minimizes the inverse susceptibility near the
ordering transition has $q_x=1/2$, but when ${\cal P}$ deviates
slightly from this critical value the $x$-component of the selected wave
vector is not exactly equal to $1/2$.  (We refer to the point 
${\cal P}={\cal P}_c$ as ``the multicritical point" because to
reach this point requires not only fixing the temperature to
be at the ordering transition, but also, as shown in Fig. \ref{INV1},
one must fix ${\cal P}={\cal P}_c$
by varying some other parameter, such as the pressure.) 
As we have seen, as the temperature is lowered into the ordered
phase when ${\cal P} \not= {\cal P}_c$, one of the
1D irreps $\Gamma_e$ or $\Gamma_o$ at $q_x \not= 1/2$ condenses,
whereas exactly at ${\cal P}={\cal P}_c$ one condenses into a phase with
$q_x=1/2$ which has only a single 2D irrep.\cite{ABH,GB05}
Accordingly, we now study the compatibility relation
which must relate the wave functions of these two phases in the limit
as we approach the multicritical point M for which ${\cal P} = {\cal P}_c$.
Experimentally, the phase transition between the phase with $q_x=1/2$ and that
having $q_x \not= 1/2$ has only been observed for $q_z=1/4$.
However, since the symmetry of the phases for $q_z=1/4$ is not different
from that for $q_z \not= 1/4$, we will leave $q_z$ as a free parameter which
we consider to be incommensurate.  Although the actual phase
transition between the $q_z=1/2$ phase and the phase with $q_z\not=1/2$
must be discontinuous, the discontinuity vanishes in the limit when 
the multicritical point M in Fig. \ref{INV1} is approached. In this limit,
one may consider the transition to be continuous, and therefore it must be
possible to express each basis function of the two irreps of the 
$q_x \not=1/2$ phase as a linear combination of the basis functions of the
2D irrep of the phase having $q_x=1/2$. We do this explicitly in order to
find the relation between the order parameters of the two phases.  This
relation will be perturbatively modified as one goes deeper into the
ordered phase.

\subsection{Wavefunctions near the Multicritical Point}

In Table \ref{SPIN} we record the wave functions allowed by symmetry
for the $q_x=1/2$ state, based on Table XVI of Ref. \onlinecite{ABH}, which are
modified in several ways.  First of all, one has to include the corrections
to the wave functions on sublattices 9-12, as described in an
erratum.\cite{ABH} Secondly, we translate all sites by $(0,0,1/2)$.
(This operation has no
effect because the induced change of phase can be absorbed into the
order  parameters.) Thirdly, we renumber the sublattices to make their
positions equal to their counterparts in Table \ref{SITES} to within a
lattice constant.  The final step was to translate sublattices through an
integer number of lattice constants, as necessary, in order to bring them 
back into the unit cell.  In this last operation sublattice  $n$
was translated through $\Delta_n$, where $\Delta_1=(0,\overline 1,0)$,
$\Delta_4=(\overline 1,0,0)$, $\Delta_9=\Delta_{10}=\Delta_{11}=(0,0,1)$,
and $\Delta_{12}=(\overline 1, \overline 1,1)$.  The result of this operation
was to introduce a  multiplicative factor $X_n=\exp[2 \pi i {\bf q} \cdot \Delta_n]$
to all components of the $n$th sublattice. Thereby we obtain the results
shown in Table \ref{SPIN}.

\begin{table} [h]
\caption{\label{SPIN} Normalized spin functions (i. e. Fourier
coefficients) within the unit cell of {\it e. g.} TbMn$_2$O$_5$ for wave
vector $(\oh,0,q)$.  Here the $r_{nx}$, $r_{ny}$ and $ir_{nz}$ are real,
the $z$'s are complex, and $\Lambda=\exp(2 \pi i q_z)$ where
$q_z$ is in rlu's.  The $x$, $y$, and $z$ components of each
Fourier vector are listed in the corresponding box.
The actual spin structure is a linear
combination, $\sigmav_1$ times the first column plus $\sigmav_2$
times the second column, where the $\sigmav$'s are complex order
parameters and the entries in each column are normalized so that
the sum of their absolute squares is unity.}
\vspace{0.2 in}
\begin{tabular}{||c||c|c|||| c || c|c ||}\hline\hline
Spin & $\sigmav_1 $ & $\sigmav_2$ & Spin  &$\sigmav_1$ & $\sigmav_2$ \\ \hline

${\bf S}({\bf q},1)$&
$\begin{array}{c} r_{1x} \\  r_{1y} \\ r_{1z}  \end{array}$ &
$\begin{array}{c} r_{2x} \\  r_{2y} \\ r_{2z}  \end{array}$  &
${\bf S}({\bf q},7)$ & 
$\begin{array}{c} z_x \\  z_y \\  -z_z  \end{array}$ &
$\begin{array}{c} z_x \\  z_y \\  z_z \end{array}$ \\ \hline

${\bf S}({\bf q},2)$&
$\begin{array}{c} r_{2x} \\  -r_{2y} \\  r_{2z}  \end{array}$ &
$\begin{array}{c} -r_{1x} \\  r_{1y} \\  -r_{1z} \end{array}$ &
${\bf S}({\bf q},8)$ &
$\begin{array}{c} z_x \\  -z_y \\  z_z  \end{array}$ &
$\begin{array}{c} -z_x \\  z_y \\  z_z \end{array}$ \\ \hline

${\bf S}({\bf q},3)$ &
$\begin{array}{c} r_{1x} \\ -r_{1y} \\  -r_{1z}  \end{array}$ &
$\begin{array}{c} -r_{2x}\\  r_{2y} \\  r_{2z}  \end{array}$  &
${\bf S}({\bf q},9)$&
$\begin{array}{c} r_{5x}\Lambda^{1/2} \\ 
r_{5y}\Lambda^{1/2} \\  r_{5z}\Lambda^{1/2}  \end{array}$ &
$\begin{array}{c} r_{6x}\Lambda^{1/2} \\  r_{6y}\Lambda^{1/2}
\\  r_{6z} \Lambda^{1/2} \end{array}$ \\ \hline

${\bf S}({\bf q},4)$ &
$\begin{array}{c} -r_{2x} \\  -r_{2y} \\ r_{2z}  \end{array}$ &
$\begin{array}{c} -r_{1x}\\  -r_{1y} \\  r_{1z} \end{array}$  &
${\bf S}({\bf q},10)$&
$\begin{array}{c} r_{6x}\Lambda^{1/2} \\
-r_{6y}\Lambda^{1/2} \\ r_{6z}\Lambda^{1/2}  \end{array}$ &
$\begin{array}{c} -r_{5x}\Lambda^{1/2} \\
r_{5y}\Lambda^{1/2} \\  -r_{5z}\Lambda^{1/2} \end{array}$ \\ \hline

${\bf S}({\bf q},5)$&
$\begin{array}{c} z_x^* \\  -z_y^* \\  -z_z^*  \end{array}$ &
$\begin{array}{c} -z_x^*\\  z_y^* \\  -z_z^* \end{array}$  &
${\bf S}({\bf q},11)$&
$\begin{array}{c} r_{5x}\Lambda^{1/2} \\
-r_{5y}\Lambda^{1/2} \\  -r_{5z}\Lambda^{1/2}  \end{array}$ &
$\begin{array}{c} -r_{6x}\Lambda^{1/2}\\ 
r_{6y}\Lambda^{1/2} \\  r_{6z}\Lambda^{1/2} \end{array}$ \\ \hline

${\bf S}({\bf q},6)$&
$\begin{array}{c} z_x^* \\  z_y^* \\  z_z^*  \end{array}$ &
$\begin{array}{c} z_x^*\\  z_y^* \\  -z_z^* \end{array}$  &
${\bf S}({\bf q},12)$&
$\begin{array}{c} -r_{6x}\Lambda^{1/2} \\
-r_{6y}/\Lambda^{1/2} \\  r_{6z}/\Lambda^{1/2}  \end{array}$ &
$\begin{array}{c} -r_{5x}\Lambda^{1/2}\\
-r_{5y}\Lambda^{1/2} \\  r_{5z}\Lambda^{1/2} \end{array}$ \\ \hline
\end{tabular}
\end{table}

Near the multicritical point M the critical spin wave
function $\Psi_{q_x=1/2}$ (for a fixed value of $q_z$ and $q_x=1/2$) is
a linear combination of $\sigma_1$ times the basis functions of the first
column of Table \ref{SPIN} plus $\sigma_2$ times the basis function of the
second column of Table \ref{SPIN}.  Alternatively, near the multicritical
point M for $q_x \not=1/2$ phase, this spin wave function can be formed
within the space in which the two 1D irreps, $\Gamma_e$ and $\Gamma_o$,
are considered degenerate for the fixed value of $q_z$.  In this limit the wave
function $\Psi_{q_x \not= 1/2}$ of the 1D irrep phase is given by a linear
combination of the basis functions associated with the four order parameters
$\sigma_s^\pm \equiv \lim_{\delta \rightarrow 0} \sigma_s 
[\pm (1/2-\delta ),0,q_z]$, where $s$ is e or o. 
These basis functions are given in Tables \ref{IIRREP} and \ref{IIIRREP}.
Equating $\Psi_{q_x=1/2}$ and $\Psi_{q_x \not= 1/2}$ gives,
with, as before, $\xi_\alpha=(-1,1,-1)$ and $\rho_\alpha=(-1,-1,1)$,
\begin{eqnarray}
&& \sigma_1  r_{1 \alpha} + \sigma_2 r_{2 \alpha} \nonumber \\
&& = \sigma_e^+ s_{\alpha ,1} + \sigma_o^+ u_{\alpha ,1}
+ \sigma_e^- \rho_\alpha s_{\alpha ,1}^*
+ \sigma_o^- \rho_\alpha u_{\alpha,1}^* 
\label{E55} \end{eqnarray}
\begin{eqnarray}
&& - \xi_\alpha [\sigma_1  r_{2 \alpha} - \sigma_2 r_{1 \alpha}] 
\nonumber \\ && = (-i \xi_\alpha)\sigma_e^+ s_{\alpha ,1}
+ (i \xi_\alpha) \sigma_o^+ u_{\alpha ,1}
\nonumber \\ && \ \ 
+ (i \xi_\alpha \rho_\alpha ) \sigma_e^- s_{\alpha,1}^*
+ ( -i \xi_\alpha \rho_\alpha) \sigma_o^- u_{\alpha,1}^*
\end{eqnarray}
\begin{eqnarray}
&& \rho_\alpha \xi_\alpha [ \sigma_1  r_{1 \alpha} - \sigma_2 r_{2 \alpha}]
\nonumber \\ &&= (i \xi_\alpha ) \sigma_e^+ s_{\alpha ,1}^*
+ (-i \xi_\alpha) \sigma_o^+ u_{\alpha ,1}^*
\nonumber \\ && \ \ 
+ (-i \rho_\alpha \xi_\alpha) \sigma_e^- s_{\alpha ,1}
+ (i \xi_\alpha \rho_\alpha) \sigma_o^- u_{\alpha,1} 
\end{eqnarray}
\begin{eqnarray}
&& \rho_\alpha [ \sigma_1  r_{2 \alpha} + \sigma_2 r_{1 \alpha}] 
\nonumber \\ &&= \sigma_e^+ s_{\alpha ,1}^* + \sigma_o^+ u_{\alpha ,1}^*
+ \sigma_e^- \rho_\alpha s_{\alpha ,1}
+ \sigma_o^- \rho_\alpha u_{\alpha,1}
\end{eqnarray}
\begin{eqnarray}
&& \xi_\alpha z_\alpha^* ( \rho_\alpha \sigma_1 + \sigma_2) 
\nonumber \\ && = \sigma_e^+ s_{\alpha ,2} + \sigma_o^+ u_{\alpha ,2}
+ \sigma_e^- \rho_\alpha s_{\alpha ,2} +
\sigma_o^- \rho_\alpha u_{\alpha ,2} 
\end{eqnarray}
\begin{eqnarray}
&& z_\alpha^* (\sigma_1 - \rho_\alpha \sigma_2) \nonumber \\
&& = (i \xi_\alpha)\sigma_e^+ s_{\alpha ,2}
+ (-i \xi_\alpha)\sigma_o^+ u_{\alpha ,2}
\nonumber \\ && \ \ 
+(-i \xi_\alpha)\sigma_e^- \rho_\alpha s_{\alpha ,2}
+ (i \xi_\alpha)\sigma_o^- \rho_\alpha u_{\alpha ,2}
\end{eqnarray}
\begin{eqnarray}
&& z_\alpha (- \rho_\alpha \sigma_1 + \sigma_2) \nonumber \\&&
= i \xi_\alpha \sigma_e^+ s_{\alpha ,2}^*
+(- i \xi_\alpha) \sigma_o^+ u_{\alpha ,2}^*
\nonumber \\ && \ \ 
+ (-i\xi_\alpha) \sigma_e^- \rho_\alpha s_{\alpha ,2}^*
+(i \xi_\alpha) \sigma_o^- \rho_\alpha u_{\alpha ,2}^*
\end{eqnarray}
\begin{eqnarray}
&& - \xi_\alpha z_\alpha ( \sigma_1 + \rho_\alpha \sigma_2) \nonumber \\
&& = \sigma_e^+ s_{\alpha ,2}^* + \sigma_o^+ u_{\alpha ,2}^*
+\sigma_e^- \rho_\alpha s_{\alpha ,2}^*
+ \sigma_o^- \rho_\alpha u_{\alpha ,2}^*
\end{eqnarray}
\begin{eqnarray}
&& (\sigma_1 r_{5,\alpha} + \sigma_2 r_{6,\alpha})\Lambda^{1/2} 
\nonumber \\ &&= \sigma_e^+ s_{\alpha, 3} + \sigma_o^+ u_{\alpha, 3}
+ \sigma_e^- \Lambda \rho_\alpha s_{\alpha, 3}^*
+ \sigma_o^- \Lambda \rho_\alpha u_{\alpha, 3}^*
\end{eqnarray}
\begin{eqnarray}
&& - \xi_\alpha (\sigma_1 r_{6,\alpha} - \sigma_2 r_{5,\alpha})\Lambda^{1/2}
\nonumber \\ && = + (- i\xi_\alpha) \sigma_e^+ s_{\alpha, 3}
+ (i \xi_\alpha) \sigma_o^+ u_{\alpha, 3} \nonumber \\ && \ \ 
+ (i \xi_\alpha) \Lambda \rho_\alpha \sigma_e^- s_{\alpha, 3}^*
+ (-i \xi_\alpha) \Lambda \rho_\alpha \sigma_o^- u_{\alpha, 3}^*
\end{eqnarray}
\begin{eqnarray}
&& \xi_\alpha \rho_\alpha (\sigma_1 r_{5,\alpha}
- \sigma_2 r_{6,\alpha})\Lambda^{1/2} \nonumber \\ && =
\Lambda (i \xi_\alpha) \sigma_e^+ s_{\alpha, 3}^*
+ \Lambda (-i \xi_\alpha) \sigma_o^+ u_{\alpha, 3}^*
\nonumber \\ && \ \ 
+ (-i \xi_\alpha) \rho_\alpha \sigma_e^- s_{\alpha, 3}
+ (i \xi_\alpha) \rho_\alpha \sigma_o^- u_{\alpha, 3}
\end{eqnarray}
\begin{eqnarray}
&& \rho_\alpha (\sigma_1 r_{6,\alpha} + \sigma_2 r_{5,\alpha})\Lambda^{1/2}
\nonumber \\ &&= \Lambda \sigma_e^+ s_{\alpha, 3}^*
+ \Lambda \sigma_o^+ u_{\alpha, 3}^* + \sigma_e^- \rho_\alpha s_{\alpha, 3}
+ \sigma_o^- \rho_\alpha u_{\alpha, 3} \ ,
\label{E66} \end{eqnarray}
where $\sigma_s^\pm \equiv \lim_{\delta \rightarrow 0} \sigma_s 
[\pm (1/2-\delta ),0,q_z]$, where $s$ is e or o. 

\begin{figure}[ht]
\begin{center}
\includegraphics[width=7.0 cm]{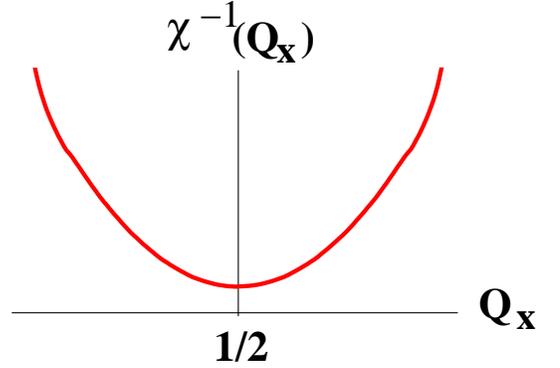}
\caption{\label{PARA1} (Color online)
The two lowest eigenvalues of $\chi^{-1}(Q_x)$ which are degenerate
for ${\cal P}={\cal P}_c$.}
\end{center}
\end{figure}

\subsection{Symmetry of the Multicritical Point}

From the above equations we expect to obtain a relation between the
order parameters of the phase with $q_x=1/2$ and that with $q_x \not= 1/2$
arbitrarily close to the multicritical point M.
Presumably, giving the values of $\sigma_e^\pm$ and $\sigma_o^\pm$
will determine the values of $\sigma_1$ and $\sigma_2$, but having
the values of $\sigma_1$ and $\sigma_2$ we can not expect to determine
the four parameters $\sigma_e^\pm$ and $\sigma_o^\pm$.  Accordingly,
we now study the basis functions for $\Gamma_e$ and $\Gamma_o$ and
show that they are related in the limit when $q_x \rightarrow 1/2$.
To see this we will analyze the behavior of the inverse
susceptibility as a function of $Q_x$, the $x$-component of the
wave vector when the temperature is just above the temperature at which
magnetic order appears and for ${\cal P}$ close to the critical
value ${\cal P}_c$ at which the minimum of the inverse susceptibility
as a function of $Q_x$
occurs for $Q_x=1/2$.  Note that the inverse susceptibility has 36
branches, each one corresponding to an eigenvalue of the inverse susceptibility
matrix.  Here we need consider only the two lowest branches of the
inverse susceptibility.  These lowest two eigenvalues
arise out of a two by two submatrix which we now analyze for
$Q_x=1/2+k_x$ and ${\cal P}={\cal P}_c +y$ for small $k_x$ and $y$.
For $y=0$ this submatrix is of the form
\begin{eqnarray}
\chiv^{-1} &=& \left[ \begin{array} {c c}
a(T-T_c) + bk_x^2 & 0 \\ 0 & a(T-T_c) + bk_x^2 \\ \end{array}
\right] \ ,
\end{eqnarray}
where $k_x=Q_x-1/2$, $a$ and $b$ are constants,  and $T_c$ is the
temperature at which order first develops. Here and below we work only
to order $k_x^2$.  This form is dictated by the
fact that the inverse susceptibility has to be two-fold degenerate, have
its minima at $k_x=0$, and the spectrum has to be independent of the
sign of $k_x$ (in view of the existence of the symmetry element $m_{bc}$).
Thus the two lowest branches in the eigenvalue
spectrum of the inverse susceptibility as a function of $Q_x$ are
as shown in Fig. \ref{PARA1}.

\begin{figure}[ht]
\begin{center}
\includegraphics[width=8.0 cm]{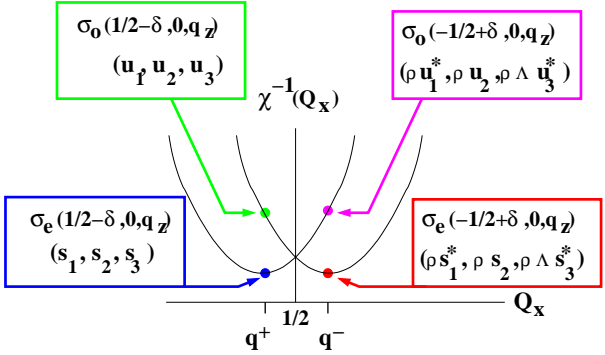}
\caption{\label{PARA2} (Color online)
The two lowest branches of eigenvalues of $\chi^{-1}(q_x)$
with their wave functions indicated. Note that the labels e and o
refer to the eigenvalues rather than the branch of the spectrum.
We assume that the wave functions at $q^+$ for irreps $\Gamma_e$ and
$\Gamma_o$ are given in terms of ${\bf s}_n$ and ${\bf u}_n$,
respectively, as listed in Tables \ref{IIRREP} and \ref{IIIRREP},
respectively.  Here ${\bf s}_n \equiv s_{\alpha,n}$, 
${\bf u}_n\equiv u_{\alpha,n}$, and $\rho \equiv \rho_\alpha$.
Then the wave functions at $q=q^-$ are obtained in terms of
those at $q=q^+$ according to Table \ref{TAMP}.  In general,
the o and e wave functions are unrelated.  However, as
${\cal P} \rightarrow {\cal P}_c$, $q^+ - q^- \rightarrow 0$
and the two parabolas come into coincidence. In this
situation the points corresponding to $\sigma_e$ and $\sigma_o$
come into coincidence.  Therefore by continuity on either the
right-hand or the left-hand parabola the $\sigma_o$ and $\sigma_e$
wave function {\it on the same parabola} become equal,
leading to Eq. (\ref{USEQ}).}
\end{center}
\end{figure}

Next consider allowed terms which are linear in $y$ but have an
unspecified dependence on $k_x$. These will give
\begin{eqnarray}
\chiv^{-1} &=& \left[ \begin{array} {c c}
\tau + bk_x^2 +c(k_x)y & d(k_x)y \\ d(k_x)^*y & \tau + bk_x^2
+ e(k_x)y \\ \end{array} \right] \ ,
\end{eqnarray}
where $c(k_x)$ and $e(k_x)$ are real and $\tau= a(T-T_c)$.
For the spectrum to be the same for both signs of $k_x$,
$c(k_x)+e(k_x)$ must be an even function of $k_x$. The term
in $[c(k_x)+e(k_x)]$
independent of $k_x$ leads to an allowed dependence of $T_c$ on $y$
and the term of order $k_x^2$ leads to an allowed dependence of the
coefficient $b$ on $y$, so, in effect, up to order $k_x^2$ we have
\begin{eqnarray}
\chiv^{-1} &=& \left[ \begin{array} {c c}
\tau + bk_x^2 + c'yk_x & d(k_x)y \\ d(k_x)^*y &
\tau + bk_x^2 - c'yk_x \\ \end{array} \right] \ ,
\end{eqnarray}
where now $\tau$ and $b$ have an allowed, but unimportant, dependence
on $y$.  Now consider the dependence of $d(k_x)$ on $k_x$.  Suppose
that $d(k_x)$ were nonzero for $k_x=0$.  This would imply that the
minimum in the inverse susceptibility occurred for $k_x=0$, but
that the eigenvalues were not degenerate.  This contradicts group
theory.  So the generic case is that $d(k_x) = \beta k_x +
{\cal O}(k_x^3)$. Then the two eigenvalues are
\begin{eqnarray}
\lambda_\pm &=& \tau+ b k_x^2 \pm y k_x \sqrt{{c'}^2 + |\beta|^2} \ .
\end{eqnarray}
This leads to two parabolic branches of the inverse susceptibility
with minima symmetrically displaced away from $Q_x=1/2$ by
an amount linear in ${\cal P}-{\cal P}_c$, as shown in Fig. \ref{PARA2}.
As shown there, the left parabola at $Q_x=q^+$ is associated with $\Gamma_e$
and is parametrized by the $s$'s and the right parabola at $Q_x=q^+$ is
associated with $\Gamma_o$ and is parametrized by the $u$'s.  The
corresponding basis functions are given explicitly in Tables \ref{IIRREP} 
and \ref{IIIRREP}.  But the basis functions for $\Gamma_o$ and $\Gamma_e$
at ${\bf q}^-$ are related, respectively, to $\Gamma_o$ and $\Gamma_e$
at ${\bf q}^+$ according to Table \ref{TAMP} and this is indicated in Fig.
\ref{PARA2}.  These eigenfunctions of the inverse susceptibility depend on
wave vector, of course. But as ${\cal P} \rightarrow {\cal P}_c$, the two parabolas
come into coincidence with their minimum at $Q_x=1/2$, and the points
governed by $\sigma_e$ and $\sigma_o$ {\it on the same parabola} approach
one another.  Then in this limit, by continuity on the {\it same} parabola
we obtain
\begin{eqnarray}
u_{\alpha,1} = \rho_\alpha s_{\alpha,1}^* , \ \ 
u_{\alpha,2} = \rho_\alpha s_{\alpha,2} , \ \
u_{\alpha,3} = \rho_\alpha \Lambda s_{\alpha,3}^* \ .
\label{USEQ} \end{eqnarray}

It should be remarked, that this multicritical point is not
a Lifshitz point.\cite{HORN} At a Lifshitz point the coefficient
of $k_x^2$ in the inverse susceptibility vanishes.  Here, in the
generic case, this coefficient is nonzero, but the coefficient
of $k_x$, which here is allowed because of the double degeneracy,
vanishes.  Furthermore, the Lifshitz point separates a regime
of CM order from that of IC order.  Here
CM order (at the paramagnetic phase boundary) only
occurs at a point (where the coefficient of $k_x$ changes sign
thereby exchanging the instabilities of the two 1D irreps).

\subsection{Compatibility Equations}

Using the relation between the ${\bf u}$'s and the ${\bf s}$'s, we see that
Eqs. (\ref{E55})-(\ref{E66}) become
\begin{eqnarray}
\sigma_1  r_{1 \alpha} + \sigma_2 r_{2 \alpha} &=&
\sigma^+ s_{\alpha ,1} + \sigma^- \rho_\alpha s_{\alpha ,1}^* \ ,
\label{E67} \end{eqnarray}
\begin{eqnarray}
\sigma_1  r_{2 \alpha} - \sigma_2 r_{1 \alpha} &=&
i \sigma^+ s_{\alpha ,1} - i \rho_\alpha \sigma^- s_{\alpha,1}^* \ ,
\label{E68} \end{eqnarray}
\begin{eqnarray}
\sigma_1  r_{1 \alpha} - \sigma_2 r_{2 \alpha}
&=& i \rho_\alpha \sigma^+ s_{\alpha ,1}^* -i \sigma^- s_{\alpha ,1} \ ,
\label{E69} \end{eqnarray}
\begin{eqnarray}
\sigma_1  r_{2 \alpha} + \sigma_2 r_{1 \alpha} &=&
\rho_\alpha \sigma^+ s_{\alpha ,1}^* + \sigma^- s_{\alpha ,1} \ ,
\label{E70} \end{eqnarray}
\begin{eqnarray}
y_\alpha^* ( \sigma_1 + \rho_\alpha \sigma_2) &=&
(\rho_\alpha \sigma^+ + \sigma^- ) s_{\alpha ,2} \ ,
\label{E71} \end{eqnarray}
\begin{eqnarray}
y_\alpha^* (\sigma_1 - \rho_\alpha \sigma_2) &=&
( i \sigma^+ -i \rho_\alpha \sigma^- ) s_{\alpha ,2} \ ,
\label{E72} \end{eqnarray}
\begin{eqnarray}
y_\alpha (\sigma_1 - \rho_\alpha \sigma_2) &=&
( - i \rho_\alpha \sigma^+ +i \sigma^- ) s_{\alpha ,2}^* \ ,
\label{E73} \end{eqnarray}
\begin{eqnarray}
y_\alpha ( \sigma_1 + \rho_\alpha \sigma_2) &=&
( - \sigma^+ - \sigma^- \rho_\alpha ) s_{\alpha ,2}^* \ ,
\label{E74} \end{eqnarray}
\begin{eqnarray}
\sigma_1 r_{5,\alpha} + \sigma_2 r_{6,\alpha} &=&
\sigma^+ s_{\alpha, 3} \Lambda^{-1/2}
\nonumber \\ && \ \ 
+ \sigma^- \Lambda^{1/2} \rho_\alpha s_{\alpha, 3}^* \ ,
\label{E75} \end{eqnarray}
\begin{eqnarray}
\sigma_1 r_{6,\alpha} - \sigma_2 r_{5,\alpha} &=&
i \sigma^+ s_{\alpha, 3}\Lambda^{-1/2}
\nonumber \\ && \ \ 
- i \Lambda^{1/2} \rho_\alpha \sigma^- s_{\alpha, 3}^* \ ,
\label{E76} \end{eqnarray}
\begin{eqnarray}
\sigma_1 r_{5,\alpha} - \sigma_2 r_{6,\alpha} &=&
i \rho_\alpha \sigma^+ s_{\alpha, 3}^*\Lambda^{1/2}
\nonumber \\ && \ \ 
-i \sigma^- s_{\alpha, 3}\Lambda^{-1/2} \ ,
\label{E77} \end{eqnarray}
\begin{eqnarray}
\sigma_1 r_{6,\alpha} + \sigma_2 r_{5,\alpha} &=&
\rho_\alpha \sigma^+ s_{\alpha, 3}^* \Lambda^{1/2}
\nonumber \\ && \ \ 
+ \sigma^- s_{\alpha, 3} \Lambda^{-1/2} \ ,
\label{E78} \end{eqnarray}
where $y_\alpha = \xi_\alpha z_\alpha$ and $\sigma^\pm =
\sigma_e^\pm + \sigma_o^\mp$. These equations are strongly
overdetermined.  Accordingly, the fact that they have a
solution is evidence that the wave functions which formed
the input to this calculation are correct.  (Indeed, in order
to arrive at a solution, it was necessary to correct an error
in the table of wave functions of Ref. 9.)
These equations have the solution for the wave functions of the
2D irrep phase in terms of those of the 1D irrep phase as
\begin{eqnarray}
r_{1, \alpha} = [e^{i \pi /4} s_{\alpha,1}
 - \rho_\alpha e^{-i \pi /4} s_{\alpha,1}^*]/\sqrt 2 \ ,
\end{eqnarray}
\begin{eqnarray}
r_{2, \alpha} = [-e^{-i \pi /4} s_{\alpha,1}
+ \rho_\alpha e^{i \pi /4} s_{\alpha,1}^*]/\sqrt 2 \ ,
\end{eqnarray}
\begin{eqnarray}
y_\alpha &=& [ - e^{i \pi /4} + \rho_\alpha e^{-i \pi/4}] s_{\alpha,2}^* \ ,
\end{eqnarray}
\begin{eqnarray}
r_{5, \alpha} = [e^{i \pi /4} \Lambda^{-1/2} s_{\alpha,3}
- \rho_\alpha e^{-i \pi /4} \Lambda^{1/2} s_{\alpha,3}^*]/\sqrt 2 \ ,
\end{eqnarray}
\begin{eqnarray}
r_{6, \alpha} = [-e^{-i \pi /4} \Lambda^{-1/2} s_{\alpha,3}
+ \rho_\alpha e^{i \pi /4} \Lambda^{1/2} s_{\alpha,3}^*]/\sqrt 2 \ .
\end{eqnarray}
The order parameters are related by
\begin{eqnarray}
\sigma^+ &=& [ e^{i \pi/4} \sigma_1 - e^{-i \pi/4} \sigma_2]/
\sqrt 2 \nonumber \\
\sigma^- &=& [ -e^{- i\pi/4} \sigma_1 + e^{i \pi/4} \sigma_2]/ \sqrt 2 \ .
\end{eqnarray}
The inverse transformation is
\begin{eqnarray}
\sigma_1 &=& [ e^{-i \pi/4} \sigma^+ - e^{i \pi/4} \sigma^-]/
\sqrt 2 \nonumber \\
\sigma_2 &=& [ -e^{i\pi/4} \sigma^+ + e^{-i \pi/4} \sigma^-]/ \sqrt 2 \ .
\end{eqnarray}

A strong check on these results is that the $r_{nx}$ and $r_{ny}$ are
real ($\rho_x=\rho_y=-1$) and $r_{nz}$ is imaginary ($\rho_z=1$),
all as required by the symmetry analysis of the CM phase.\cite{ABH}
 
These results show how the order parameters of the 2D irrep are
related to the order parameters of the 1D irreps. One should also
note that by continuity, if the IC phase has a spontaneous 
polarization as $q_z \rightarrow 1/4$, the CM phase should
also have one, and vice versa.  This is ensured by the fact that
\begin{eqnarray}
|\sigmav_1|^2 - |\sigmav_2|^2 &=& i[(\sigmav_e^+ + \sigmav_o^-)( \sigmav_e^-
+ \sigmav_o^+)^* \nonumber \\ && \ \
- (\sigmav_e^+ + \sigmav_o^-)^* (\sigmav_e^- + \sigmav_o^+) ] \ .
\end{eqnarray}
Now we only keep terms which conserve wave vector when we go away from
$q_x=1/2$, in which case
\begin{eqnarray}
|\sigmav_1|^2 - |\sigmav_2|^2 &=& i[\sigmav_e^+\sigma_o^{+*}
- \sigmav_o^+ \sigmav_e^{+*}\nonumber \\ && \ \
+ \sigmav_e^{-*} \sigmav_o^- - \sigmav_o^{-*} \sigmav_e^- ] \ .
\end{eqnarray}

Thus the ME interaction of Eq. (\ref{E53}) goes smoothly into
the ME interaction in the CM state\cite{ABH,HAE,FOCUS}
\begin{eqnarray}
V_{\rm int} &=& r [ |\sigmav_1|^2 - |\sigmav_2|^2] P_b \ .
\end{eqnarray}
 
\section{Conclusion}
We have performed a representation analysis of the magnetic order
for the IC phase of the RMn$_2$O$_5$ series by including
inversion symmetry, thereby reducing by about half the number of
degrees of freedom allowed for magnetic ordering.  Our results
emphasize that a full inclusion of inversion symmetry is necessary to
determine the magnetic structure and associated order parameters, not
only in multiferroics, but also in a wide range of magnetic materials.
We have also determined the
physically important order parameters and have analyzed the transformation
properties which they inherit from the wave functions.
Using these symmetry properties we have analyzed the magnetoelectric
interaction responsible for the simultaneous magnetic and dielectric
phase transitions.  The lowest order magnetoelectric interaction,
which is bilinear in the magnetic order parameters,
explains the observed direction of the spontaneous polarization.
We have shown that higher order and {\it Umklapp} magnetoelectric
interactions (which are quartic in the spin variables)
can induce nonzero values for all components of the
spontaneous polarization, but since the order parameters are
small in the relevant phases and since microscopic mechanisms tend to
involve terms quadratic in the spin variables, these anomalous
components to the spontaneous polarization may be very difficult
to observe.  We have also explicitly obtained
the compatibility relations for the transition between the IC
phase and the CM phase (or more generally the phase where the
$x$-component of wave vector is locked to its CM value).

\noindent {\bf Acknowledgements}
AA and OEW acknowledge support from the Israel Science Foundation.
MK acknowledges support by the Swiss National Science Foundation under
contract No. PP002-102831.


\begin{thebibliography} {99}
\bibitem{LTIC}
The very low temperature IC phases may be long wave length
IC phases.  See Refs. 11 and 12.
\bibitem{FN}
For a phase directly connected to the paramagnetic phase via a
continuous transition we ignore the possibility of accidentally degenerate
irreps, so only one irrep can appear.  If a phase is reached by
passing through two continuous phase transitions, then we
similarly assume the presence of two irreps.
\bibitem{FN3}
Throughout the paper $q_x$ is assumed to be near or at 1/2 and
$q_z$ near or at 1/4.
\bibitem{FN2}
The factor $2 \pi$ occurs because positions are measured in lattice
constants and wave vectors are in rlu's.
\bibitem{QFN}
Here we assume that the e and o order parameters condense at the same
wave vector.  This assumption is justified in Refs. \onlinecite{PRB},
\onlinecite{FOCUS}, and \onlinecite{HAE}.
\end{thebibliography}
\end{document}